\documentclass[aps,reprint,floatfix]{revtex4-1}
\usepackage{amsmath,amssymb,graphicx,mathtools}
\usepackage{float}

\usepackage[T1]{fontenc}

\usepackage{colortbl}

\makeatletter
\DeclareRobustCommand{\cev}[1]{%
  \mathpalette\do@cev{#1}%
}

\makeatother

\newcounter{theoremcounter}
\stepcounter{theoremcounter}

\newcommand{\bs}{\boldsymbol}
\newcommand{\bra}[1]{\left\langle #1\right|}
\newcommand{\ket}[1]{\left|#1\right\rangle}
\newcommand{\braket}[2]{\left\langle #1|#2\right\rangle}

\newcommand{\ketbra}[2]{\ket{#1}\bra{#2}}

\DeclareMathOperator{\Tr}{Tr}

\newcommand{\Mod}[1]{\ \mathrm{mod}\ #1}
\newcommand{\bsmc}[1]{\bs{\mathcal #1}}

\makeatletter
\newcommand{\vast}{\bBigg@{4}}
\newcommand{\Vast}{\bBigg@{5}}
\makeatother

\begin{document}

\title{Improved Simulation of Quantum Circuits by Fewer Gaussian Eliminations}
\author{Lucas Kocia}
\affiliation{Sandia National Laboratories, Livermore, California 94550, U.S.A.}
\author{Mohan Sarovar}
\affiliation{Sandia National Laboratories, Livermore, California 94550, U.S.A.}
\begin{abstract}
  We show that the cost of strong simulation of quantum circuits using \(t\) \(T\) gate magic states exhibits non-trivial reductions on its upper bound for \(t=1\), \(t=2\), \(t=3\), and \(t=6\) with odd-prime-qudits. This agrees with previous numerical bounds found for qubits. We define simulation cost by the number of terms that require Gaussian elimination of a \(t \times t\) matrix and so capture the cost of simulation methods that proceed by computing stabilizer inner products or evaluating quadratic Gauss sums. Prior numerical searchs for qubits were unable to converge beyond \(t=7\). We effectively increase the space searched for these non-trivial reductions by \(>10^{10^4}\) and extend the bounds to \(t=14\) for qutrits. This is accomplished by using the Wigner-Weyl-Moyal formalism to algebraically find bounds instead of relying on numerics. We find a new reduction in the upper bound from the \(12\)-qutrit magic state of \({3^{\sim 0.469t}}\), which improves on the bound obtained from the \(6\)-qutrit magic state of \({3^{\sim 0.482t}}\).
\end{abstract}
\maketitle

\emph{Strong quantum simulation} is the task of calculating probabilities of arbitrary output strings of universal quantum circuits. This task is \(\#P\)-hard~\cite{Huang18_2} and therefore any classical algorithm that improves on prior attempts can likely only lower the exponential coefficient in the cost of strong quantum simulation. However, such improvements are important for (i) simulating small-scale (NISQ~\cite{Preskill18}) quantum devices, and (ii) understanding the limits of classical simulation of quantum computers and the practical onset of quantum advantage~\cite{Harrow17}.

Universal quantum computations can be written in terms of \(k\)-tensored \(T\) gate magic states \(\ket T ^{\otimes k}\) with \((n-k)\) computational states, that are then acted on by Clifford gates \(\hat U_C\) and partially traced over to obtain a marginal over any qudit~\cite{Bravyi05}:
\begin{equation}
  \label{eq:probofapp}
  P_k = \Tr \left[ \hat \Pi \hat \rho \right],
\end{equation}
where \(\hat \rho = \ketbra{\Psi}{\Psi}\) for \(\ket \Psi = \hat U_C \ket{T}^{\otimes k}\ket{0}^{\otimes (n-k)}\) and \(\hat \Pi\) is a projector onto a Pauli operator eigenstate. A similar statement can be made by reformulating universal quantum computation in terms of Pauli-based-computation~\cite{Bravyi16_2}. In this equation, the \(T\) gate magic state, which for qubits is \(\ket T^{\otimes k} = \frac{1}{\sqrt 2}( \ket 0 + e^{\pi i/4} \ket 1)^{\otimes k}\), is a resource state that extends the Clifford classical subtheory to quantum universality in the limit of \(k \rightarrow \infty\)~\cite{Bravyi05}. We expect the cost of computing \(P_k\) to grow exponentially with \(k\) since this parameter dictates the degree of non-classicality of the circuit.

Stabilizer states \(\ket{\phi_i}\) are eigenvalue \(+1\) eigenstates of an Abelian subgroup of the Pauli group and they form an over-complete basis in Hilbert space. Since Cliffords simply permute stabilizer states, Eq.~\ref{eq:probofapp} can be rewritten in terms of only stabilizer states by simply expanding the \(T\) gate magic states in terms of their stabilizer decomposition: \(\ket \Psi = \sum_{i=1}^{m_k} c_i \ket{\phi_i}\) for some \(c_i \in \mathbb C\) (since \(\ket 0\) is a stabilizer state). Since Pauli projections take stabilizer states to stabilizer states, this expansion means Eq.~\ref{eq:probofapp} can be expressed as a linear combination of inner products of stabilizer states, \(P_k = \sum_{i,j=1}^{m_k} c_i c^*_j \braket{\phi_j}{\phi'_i}\), where \(\ket {\phi'_i} = \hat \Pi \ket {\phi_i}\). The inner product of two \(n\)-qudit stabilizer states, \(\braket{\phi_i}{\phi_j}\), is governed by Gaussian elimination and therefore scales as \(\mathcal O (n^3)\). Thus, calculating \(P_k\) scales as \(\mathcal O(m^2_k k^3)\). This scaling can be improved to \(O(m_k k^3)\) by instead using an estimation technique that computes inner products between \(\phi_i\) and random stabilizer states from a uniform distribution~\cite{Bravyi16_1}.

Let \(\chi_k\) be the \emph{stabilizer rank} of the \(T\) gate magic state \(\ket T^{\otimes k}\)---the minimal number of states required in a stabilizer state decomposition of \(\ket T^{\otimes k}\). Therefore, the smallest that \(m_k\) can be is \(\chi_k\) and determining its scaling with \(k\) is crucial for understanding the optimal \(\mathcal O (\chi_k k^3)\) cost of classically computing \(P_k\). Although the \(T\) gate magic state is not the unique resource state, we focus on it in this study because it is postulated that its stabilizer rank \(\chi_k\) grows slowest with \(k\)~\cite{Bravyi16_2}.

The property that the tensor product of two stabilizer states is a stabilizer state implies a \emph{trivial tensor bound} on the stabilizer rank for all integer powers of a state: \(\chi_t \le (\chi_k)^{t/k} \) where \(t\) is a multiple of \(k\). However, it is possible that the actual stabilizer rank $\chi_t$ is strictly less than this trivial bound. If so, this implies that $\ket{\Psi}^{\otimes t'}$ has a more efficient stabilizer decomposition, for $t'$ any multiple of $t$. Therefore, it is important to identify such reductions in rank over the trivial bound, a problem we tackle in this paper.

Prior searches for these improved tensor bounds for the qubit \(T\) gate magic state have relied on numerical Monte Carlo searchs of the stabilizer space (Glauber dynamics)~\cite{Bravyi16_2}. The results can be summarized in terms of four values: \(\chi_1 = 2\), \(\chi_2 = 2\), \(\chi_3 = 3\), and \(\chi_6 \le 7\). This last bound is conjectured to be tight~\cite{Howard18}. From these four data points and their tensor upper bounds, one can surmise that \(\chi_4 \le (\chi_2)^2 = 4\), \(\chi_5 \le \chi_3 \chi_2 = 6\), and \(\chi_7 \le (\chi_2)^2 \chi_3 = 12\). These bounds are likewise conjectured to be tight and numerical searches support this claim~\cite{Howard18}. The tensor bound implies the following upper bounds on the \(T\) gate stabilizer rank: \(\chi_t \le (\chi_1)^t = 2^t\) for arbitrary \(t\), \(\chi_t \le (\chi_2)^{t/2} = 2^{0.5 t}\) for even \(t\), \(\chi_t \le (\chi_3)^{t/3} = 2^{\sim 0.53 t}\) for \(t\) a multiple of \(3\), and \(\chi_t \le (\chi_6)^{t/6} = 2^{\sim 0.47 t}\) for \(t\) a multiple of \(6\). These applications of the trivial tensor bound tell us about the asymptotic scaling of the strong simulation cost of Eq.~\ref{eq:probofapp} and it is clear that the last bound provides the most favorable such scaling.

To find a better asymptotic scaling requires reaching larger \(t\). Unfortunately, the number of stabilizer states grows as \(2^{(1/2+o(1))t^2}\)~\cite{Aaronson04} and the stabilizer rank grows at least linearly with \(t\), therefore any numerical search must contend with a prohibitive search space of size \({>2^{(1/2+o(1))t^3}}\). Monte Carlo stops converging appreciably on current hardware at \(t>7\). Therefore, a non-numerical method is especially desirable.

In this direction, we previously showed that odd-prime-\(d\) dimensional qudit \(T\) gate magic states have the same stabilizer rank for \(t=1\) and \(t=2\) as has been found for qubits up to the exponential base factor---\(2^{\alpha t} \leftrightarrow d^{\alpha t}\), i.e. \((\chi_1)^t = d^{t}\) and \((\chi_2)^t = d^{0.5 t}\)~\cite{Kocia18_2}. In fact, we proved that stabilizer decompositions that achieve these stabilizer ranks for \(t=1\) and \(t=2\) have a one-to-one correspondence with the quadratic Gauss sums that decompose the \(T\) gate magic state's discrete Wigner function, which are operationally and cost-wise equivalent to the stabilizer rank. Quadratic Gauss sums are the discrete analogue of Gaussian integrals: \(\sum_{\bs x \in (\mathbb Z/ d \mathbb Z)^n} \exp [\frac{2 \pi i}{d} (\bs x^T \bsmc A \bs x + \bs \beta \cdot \bs x)]\), where \(\bsmc A \in \mathbb Z^{n \times n}\) and \(\bs \beta \in \mathbb Z^n\). Finding the minimum number of quadratic Gauss sums can be accomplished with an algebraic approach and so can be extended to higher numbers of qudits. Here we push this analysis further and find that reductions in the number of quadratic Gauss sums over the trivial tensor bound also exist for \(t=3\) and \(t=6\) as they did for qubits, the latter producing a scaling bound of \(3^{\sim 0.482 t}\) for qutrits (for \(t\) a multiple of \(6\)). Unlike the numerical approach, we are able to push far past \(t=7\) and extend our search to \(t=14\). We find that the upper bound cannot be improved over the trivial tensor bound until \(t=12\), where the new rank produces an improved scaling bound of \(<3^{\sim 0.469 t}\) for qutrits (for \(t\) a multiple of \(12\)).

In the following, we will first introduce the Wigner-Weyl-Moyal (WWM) formalism that forms the basis of our approach and explain why its quadratic Gauss sums are operationally equivalent to stabilizer state inner products. We then sketch how to use the WWM formalism to algebraically determine the bound on the minimal number of quadratic Gauss sums necessary to evaluate the Wigner function of the \(T\) gate magic state. This is followed by our main results for \(t=3\), \(t=6\) and \(t=12\).

\paragraph*{\bf{\emph{The WWM Formalism}}} Instead of considering the magic state in terms of vectors in Hilbert space, we consider a kernel (or quasi-probability) representation; given a complete set of Hilbert-Schmidt orthogonal operators \(\hat R(\bs x)\), indexed by \(\bs x \equiv (\bs x_p, \bs x_q) \in ((\mathbb Z/ d \mathbb Z)^{n})^2\), any operator \(\hat A \in \mathcal B(\mathcal (\mathbb C^d)^n)\) can be represented as
\begin{equation*}
  \hat A = d^{-1} \sum_{\substack{\bs x \in\\ (\mathbb Z/ d \mathbb Z)^{2n}}} \Tr (\hat R(\bs x) \hat A) \hat R(\bs x) \equiv \sum_{\bs x} A(\bs x) \hat R(\bs x).
\end{equation*}

In particular, we consider the odd-prime \(d\)-dimensional Weyl operators~\cite{Wootters87,Rivas99, Rivas00},
\begin{equation*}
  \hat R(\bs x) = d^{-n} \sum_{\substack{\bs y_p, \bs y_q \in\\(\mathbb Z/ d \mathbb Z)^{n}}} e^{\frac{2 \pi i}{d} (\bs y_p \cdot \bs x_q - \bs y_q \bs x_p -\frac{1}{2} \bs y_p \cdot \bs y_q)} \hat Z^{\bs y_p} \hat X^{\bs y_q},
\end{equation*}
where \(\hat X\) and \(\hat Z\) are \(d\)-dimensional generalized Pauli operators~\cite{Weyl32}. In the case when \(\hat A = \hat \rho\) is a quantum state, \(A(\bs x) = \rho(\bs x)\) is called a Wigner function. Otherwise, it is called a Weyl symbol. \(\hat R(\bs x)\) are Hermitian, self-inverse and unitary and so the coefficients \(\rho(\bs x)\) are real-valued. This representation is particularly simple for the Clifford subtheory: the Wigner function of states \(\rho(\bs x)\) are non-negative if and only if they are stabilizer states~\cite{Gross06,Gross07} and the Weyl symbol of Clifford gates \(U_C(\bs x)\) are symplectic positive maps that can be described as affine transformations:~\cite{Gross06,Kocia17}: \(\bs x' \equiv \left( \bs x'_p, \bs x'_q \right)^T = \bsmc M_C \left( \bs x_p, \bs x_q \right)^T + \bs v_C\). The form of the \(\bsmc M_C\) and \(\bs v_C\) for the Clifford gates are given in~\cite{Kocia17}.

In the WWM formalism, Eq.~\ref{eq:probofapp} becomes 
\begin{equation}
  P_k = \sum_{\bs x \in D} \left[ \prod_{i=1}^k \rho_{T}(\bs x_i) \prod_{j=k+1}^{n} \delta(x_{q_j})\right],
  \label{eq:applicationtrace}
\end{equation}
for
\begin{equation}
  D = \left\{\bs x \bigg| \left(\bsmc M_C^{-1} \bs x + \bs v \right)_{n+1} \Mod d^h =0\right\}
  \label{eq:applicationtracerestriction}
\end{equation}
for some \(h \in \mathbb Z^+\) and \((\bs x)_i\) is the \(i\)th element of \(\bs x\). The Clifford sequence \(\hat U_C\) changes the restriction of the domain of the sum from \(x_{n+1} \equiv x_{q_1} = 0\) to \(D\).

We showed previously~\cite{Kocia18_2} that 
\begin{equation}
\rho_{T^{\otimes k}}(\bs x) = \prod_{i=1}^k \rho_T(\bs x_i) = \sum_{\bs y_q \in (\mathbb Z/ d \mathbb Z)^k} e^{\frac{2 \pi i}{ d^h} P(\bs y_{q}, \bs x)},
\label{eq:intermediatesum}
\end{equation}
for \(P\) a polynomial in \(\bs y_{q}\) and \(\bs x\) over \(\mathbb Z\), for \(p\)-odd-prime qudits. We refer to \(\bs y_q\) as intermediate variables in order to distinguish them from \(\bs x \equiv (\bs x_p, \bs x_q)\), which are the final variables at which the Wigner function is evaluated.

\begin{table*}[t]
  \begin{tabular}{p{30pt}|p{30pt}|p{30pt}|p{30pt}|p{30pt}|p{30pt}|p{30pt}|p{30pt}|p{30pt}|p{30pt}|p{30pt}|p{30pt}|p{30pt}|p{30pt}|p{30pt}}
    \hline
    \(k\) & $1$ & $2$ & $3$ & $4$ & $5$ & $6$ & $7$ & $8$ & $9$ & $10$ & $11$ & $12$ & $13$ & $14$ \\
    \hline \hline
    \multicolumn{8}{l}{qubit:} \\
    \hline
    \(\chi_k\) & $2$ & $2$ & $3$ & $4$ & $6$ & $7$ & $12$ & \multicolumn{7}{c}{\cellcolor[gray]{0.8}\emph{inaccessible to Monte Carlo}}\\
    \hline
    \(\chi_k^{t/k}\) & {$2^t$}& {$2^{0.5 t}$}& {$2^{\sim 0.528t}$}& & & {${2^{\sim 0.468t}}$} & & \multicolumn{7}{c}{\cellcolor[gray]{0.8}}\\
    \hline
    \multicolumn{8}{l}{qutrit:} \\
    \hline
    \(\chi_k\) & $3$ & $3$ & $8?$ & \multicolumn{11}{c}{\cellcolor[gray]{0.8}\emph{inaccessible to Monte Carlo}}\\
    \hline
    \(\xi_k\) & $3$& $3$ & $8$ & $9$ & $24$ & $24$ & $\le 72$ & $72$ & $\le 216$ & $216$ & {$\le 486$} & {$486$} & $\le 1458$& $1458$\\
    \hline
    \(\xi_k^{t/k}\) & {$3^t$}& {$3^{0.5 t}$}& {$3^{\sim 0.631t}$}& & & {$3^{\sim 0.482t}$} & & & & & {$3^{\sim 0.512 t}$} & {$3^{\sim 0.469 t}$} & & \\
    \hline
  \end{tabular}
  \caption{Upper bound of qubit and qutrit \(T\) gate magic state stabilizer ranks \(\chi_k\) are tabulated and compared to qutrit quadratic Gauss sum ranks \(\xi_k\) along with their tensor upper bounds (\(\chi_k^{t/k}\) and \(\xi_k^{t/k}\) respectively). The reductions in the qubit scaling for \(k=1\), \(k=2\), \(k=3\) and \(k=6\) are observed for qutrits as well. Moreover, a further reduction is observed for qutrits for \(k=12\), a result beyond the reach of Monte Carlo numerical search. \((\chi_k)^{t/k}\) is only listed at the \(k\) values at which there is a reduction over the trivial tensor bound.}
  \label{tab:Gausssumresults}
\end{table*}

For instance, we found that the Wigner function of the two-qutrit tensored \(T\) gate magic state, \(\ket{ T}^{\otimes 2} = (\ket 0 + e^{\frac{2 \pi i}{9}} \ket 1 + e^{-\frac{2 \pi i}{9}} \ket 2)^{\otimes 2}\), can be written~\cite{Kocia18_2},
  \begin{eqnarray}
    \label{eq:twoqutritpi8gate_2}
    \frac{1}{3^2} \sum_{y_{q_1} \in \mathbb Z/ 3 \mathbb Z} \exp \left\{ \frac{2 \pi i}{3^2} \left[ 8 x_{q_1}^3 + 7 y_{q_1}^3 \right] \right\} \mathcal A_2(y_{q_1}, \bs x).
  \end{eqnarray}
where 
\begin{eqnarray*}
  \mathcal A_2(y_{q_1}, \bs x)&=& \sum_{y_{q_2} \in \mathbb Z/ 3 \mathbb Z} e^{ \frac{2 \pi i}{3} P(y_{q_1}, y_{q_2}, \bs x)},
\end{eqnarray*}
for \(P(y_{q_1}, y_{q_2}, \bs x)\) a polynomial over \(\mathbb Z\) that is quadratic in \(y_{q_2}\). See the Appendix~\ref{app:twoqutritmagicstate} for the full form of \(\mathcal A_2\).

Therefore, Eq.~\ref{eq:twoqutritpi8gate_2} is a Wigner function that is a linear combination of three terms indexed by \(y_{q_1}\), each of which is a quadratic Gauss sum over \(y_{q_2}\). Quadratic Gauss sums require \(\mathcal O(k^3)\) computations to evaluate for \(k\) qudits; their absolute value depends on the determinant of their covariance matrix and so they are governed by the cost of Gaussian elimination of a matrix of size \(k\times k\) with entries in \(\mathbb  Z/ d \mathbb Z\)~\cite{Kocia18_2}. Importantly, in order to calculate stabilizer state inner products, a ``tableau'' matrix which has the same properties must also undergo Gaussian elimination~\cite{Aaronson04}. The cost of calculating the phase of the quadratic Gauss sum is similarly equivalent to the cost of evaluating the phase of stabilizer state inner products. WWM quadratic Gauss sums are thus operationally equivalent to Hilbert space stabilizer state inner products. 

As a result, we proceed to determine the cost of evaluating Eq.~\ref{eq:applicationtrace} in terms of the number of quadratic Gauss sums in its sum. We previously showed that to find this number it is sufficient to just determine the number of quadratic Gauss sums necessary to evaluate \(\rho_{T^{\otimes k}}(\bs x)\) for fixed \(\bs x\)~\cite{Kocia18_2}\footnote{This is true as long as \(\bs x_{q}\) and \(\bs y_{q}\) have similar polynomials for \(\bs y_q\) and \(\bs x_q\) fixed, respectively. i.e. if \(x_{q_i}\) is cubic, then so is \(y_{q_i}\) (and vice-versa).}. We define the minimum of this number \(\xi_k\), and call it the \emph{quadratic Gauss sum rank}. More precisely, the overall cost of evaluating Eq.~\ref{eq:applicationtrace} scales as \(\mathcal O(\xi_k)\)~\cite{Kocia18_2}. Since the product of two Wigner functions on separate qudits is also a Wigner function, \(\xi_k\) satisfies the same trivial tensor bound property as \(\chi_k\): \(\xi_t \le (\xi_k)^{t/k}\) for \(t\) a multiple of \(k\). Therefore, calculating Eq.~\ref{eq:applicationtrace} using the WWM formalism scales as \(\mathcal O(\xi_k k^3)\), similarly to how using stabilizer state inner products scales as \(\mathcal O(\chi_k k^3)\).

Here we employ this approach to operationally define the cost of evaluating \(P_k\) in Eq.~\ref{eq:applicationtrace}. In particular, this paper examines the odd-prime \(d\)-dimensional qudit \(\xi_k\) for \(k>2\). We focus on the smallest such qudit: the qutrit (\(d=3\)).

\paragraph*{\bf{\emph{Results}}} In much the same way that stabilizer decompositions of a state are generally non-unique, decompositions of a Wigner function in terms of quadratic Gauss sums are also generally non-unique. In the WWM formalism, this freedom is due to the invariance of the discrete sum in Eq.~\ref{eq:intermediatesum} under linear transformations of its variables \(\bs y_q\) as these lie on a finite odd-prime field \(\mathbb Z/ p \mathbb Z\). This is true for both the intermediate variables \(\bs y_q\) and the final phase space variables \(\bs x\) when the full trace is taken in Eq.~\ref{eq:applicationtrace}, despite the additional restriction in the domain (see Appendix~\ref{app:CNOTfreedom} for a more detailed discussion). We use this freedom to algebraically lower the number of quadratic Gauss sums that are naively obtained from the trivial tensor bound for the Wigner function of higher tensor powers of the \(T\) gate magic state. We find that products of linear transformations corresponding to the Clifford controlled-not \(C_{i,j}\) gate between qudit \(i\) and \(j\), \(\bsmc M_{C_{i,j}}: (x_{p_i}, x_{p_j}, x_{q_i}, x_{q_j}) \rightarrow (x_{p_i}, x_{p_j}-x_{p_i} \Mod d, x_{q_i}+x_{q_j} \Mod d, x_{q_j})\), are sufficient for this purpose.

As mentioned earlier, the one- and two-qutrit Wigner functions of the \(T\) gate magic states can be written in terms of three quadratic Gauss sums (i.e. see Eq.~\ref{eq:twoqutritpi8gate_2})~\cite{Kocia18_2}. Hence, \(\xi_1 = \xi_2 = 3\), which leads to tensor bounds of \(\xi_t \le (\xi_1)^t = 3^t\) and \(\xi_t \le (\xi_2)^{t/2} = t^{0.5 t}\), for even \(t\).

The trivial tensor bound indicates that \(\xi_3 \le 9\). After transformation by a \(C_{1,2}^2\), \(C_{1,3}^2\) and \(C_{2,3}\) (the overall transformation we call \(C_3\)), we find that the three-qutrit \(T\) gate magic state can be written as:
  \begin{eqnarray}
    \label{eq:summarythreequtrit}
    &&\rho_{T^{\otimes 3}}({\bsmc M}_{C_3} \bs x)\\
    &=& \sum_{\substack{y_{q_1}, y_{q_2}\\ \in \mathbb Z/ 3^2 \mathbb Z}} \exp \left[ \frac{2 \pi i}{9} \left( 7 y_{q_1}^3 + 8 x_{q_1}^3 \right) \right] \mathcal A_3 (y_{q_1}, y_{q_2}, \bs x) \nonumber\\
    && \qquad \times \left[ \delta(\neg (y_{q_1}-x_{q_1})) + \delta(y_{q_1}-x_{q_1})\delta(\Delta)  \right],\nonumber
  \end{eqnarray}
where \(\delta(\neg \alpha) = \delta((\alpha^{p-1}-1)^{p-1})\) is logical negation. Logical negation of an argument \(\alpha\) for prime \(p\) is simply \(\alpha^{p-1} \Mod p\): if \(x \ne 0\) then \(x^{p-1} \Mod p = 0\) and if \(x = 0\) then \((x-1)^{p-1} \Mod p = 1\). \(\mathcal A_3\) is a quadratic Gauss sum and \(\Delta\) is a the linear coefficient of this sum (see Appendix~\ref{app:threequtritmagicstate} for their explicit form).

Eq.~\ref{eq:summarythreequtrit} is a linear combination of nine quadratic Gauss sums indexed by \(y_{q_1}\) and \(y_{q_2}\). However, the additional Kronecker delta functions explicitly express the contrapositive of the condition that these quadratic Gauss sums are zero at \(y_{q_1}=x_{q_1}\) and \(\Delta \in \{1, 2\}\). Moreover, the delta function terms are disjoint; given any \((\bs x_p, \bs x_q)\) and \(\bs y_{q}\), only one term can be non-zero. However, given \(\bs x\), all the terms are zero for at least one value of \((\bs y_{q_1}, \bs y_{q_2})\) in the sum, thereby reducing the number of quadratic Gauss sums.

Hence, the Wigner function of three tensored qutrit magic states can be expressed in terms of only \(\xi_3 = 8\) non-zero quadratic Gauss sums. Extrapolating to higher \(t\) counts using the tensor bound, this result shows that \(\xi_3^{t/3} = 3^{\frac{\log 8}{3 \log 3}t} = 3^{\sim 0.63 t}\) quadratic Gauss sums can represent \(t\) magic states, for \(t\) a multiple of \(3\). We also find numerical evidence that \(\chi_3=8\) (see Appendix~\ref{app:numerics}) from running the same Monte Carlo search algorithm as~\cite{Bravyi16_2} further estabilishing evidence that \(chi_k = \xi_k\) for \(k=1\), \(2\), and \(3\).

A similar reduction from the trivial tensor bound can be found for the six qutrit \(T\) gate magic state (see Appendix~\ref{app:sixqutritmagicstate}) from some quadratic Gauss sums evaluating to zero. However, the six-qutrit case undergoes a further reduction due to two sets of quadratic Gauss sums evaluating to the same value. These sets are indexed the cubic intermediate variable \(y_{q_3}\), for \(y_{q_1}\) fixed. In the worst case over Clifford gates \(\hat U_C\) in Eq.~\ref{eq:probofapp}, only this last reduction occurs and so the Wigner function consists of \(\xi_6 = 24 (=3^3-3)\) quadratic Gauss sums. This leads to a trivial tensor bound of \(\xi_t \le (\xi_6)^{t/6} = 3^{\sim 0.482 t}\) for \(t\) a multiple of \(6\).

Lastly, the twelve qutrit \(T\) gate magic state also exhibits a reduction (see Appendix~\ref{app:twelvequtritmagicstate}). Similarly to the six-qutrit case, in the worst case over Clifford gates \(\hat U_C\), two sets of quadratic Gauss sums evaluate to the same sum and are indexed by two values of the intermediate cubic variable \(y_{q_6}\). However, unlike for the six-qutrit case, this condition holds for more indexing variables than would be proportionally expected: \(y_{q_1}\), \(\ldots\), \(y_{q_4}\). Therefore, in the worst-case this Wigner function consists of \(\xi_{12} = 3^4\times 6 = 486\) quadratic Gauss sums. This leads to a trivial tensor bound of \(\xi_t \le (\xi_{12})^{t/12} = 3^{\sim 0.469 t}\), for \(t\) a multiple of \(12\).

Results up to \(t=14\), including the results discussed above, are tabulated in Table~\ref{tab:Gausssumresults}. 

\begin{figure}[ht]
  \includegraphics[scale=0.3]{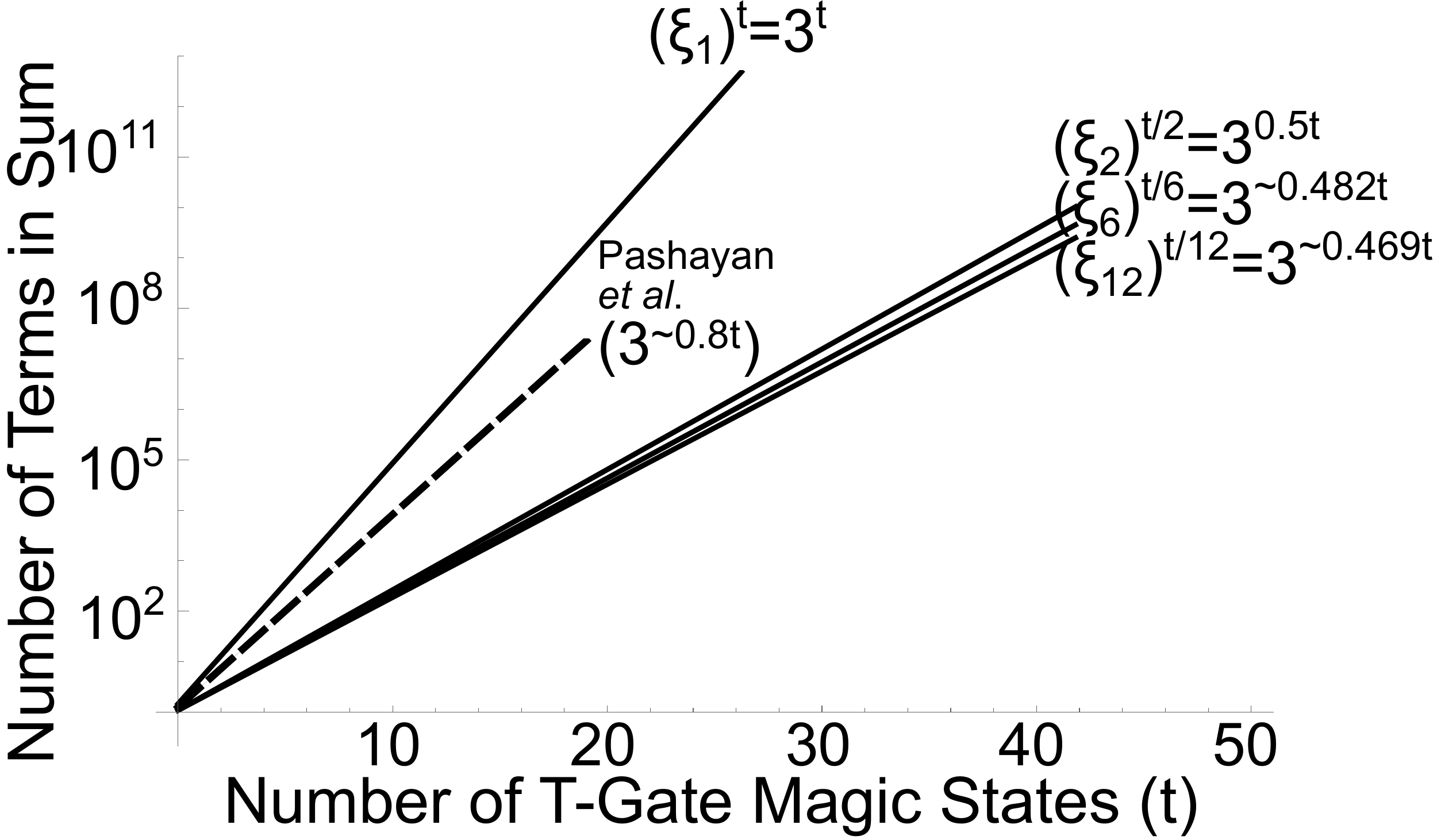}
  \caption{Logarithm of the worst-case number of terms required to evaluate \(P_k\) for qutrits in Eq.~\ref{eq:applicationtrace} for a Monte Carlo method based on Wigner negativity~\cite{Pashayan15} (dashed curve) compared to the qutrit trivial tensor bound from \((\xi_1)^t\), \((\xi_2)^{t/2}\), \((\xi_6)^{t/6}\) and \((\xi_{12})^{t/12}\) (solid curves).}
  \label{fig:pi8gatenumerics}
\end{figure}

The trivial tensor bounds set the cost of classical strong simulation of \(P_k\), and we can compare this cost to that of existing simulation methods. In Figure~\ref{fig:pi8gatenumerics}, we compare these bounds to the cost of a Monte Carlo numerical method based on qutrit Wigner function sampling~\cite{Pashayan15}~\footnote{The Monte Carlo was evaluated to precision \((P_{\text{sampled}} - P) <10^{-2}\) with \(95\%\) confidence. Only the exponential factor of the Monte Carlo method's cost is shown in the figure.} Note that the direct evaluation of \(P_k\) using the WWM formalism is an explicit algorithm (see Table \(1\) in~\cite{Kocia18_2}) that saturates the bounds shown in Figure~\ref{fig:pi8gatenumerics}. We find that the WWM algorithm provides an exponential improvement over existing methods. For example, \(P_{35}\) can be simulated exactly by evaluating \(10^8\) quadratic Gauss sums whereas for \(10^8\) samples the Monte Carlo method in~\cite{Pashayan15} only allows for evaluating up to \(P_{10}\).

\paragraph*{\bf{\emph{Discussion}}} Using the same argument as in~\cite{Aaronson04}, and the fact that the \(n\)-qutrit stabilizer group has size \(3^n\)~\cite{Gheorghiu14}, we find that the number of pure stabilizer states on \(n\) qutrits is
\begin{equation*}
  3^n \prod^{n-1}_{k=0} (7^n/2^n-2^k) / \prod^{n-1}_{k=0} (3^n-2^k) = 3^{(1/2 + o(1))n^2}.
\end{equation*}
This space grows faster than the \(2^{(1/2 + o(1))n^2}\) \(n\)-qubit stabilizer subspace~\cite{Aaronson04} and so would be more difficult to search using Monte Carlo techniques. Nevertheless, with the algebraic approach presented here, we are able to comfortably bound well past \(t=7\) to \(t=14\), an increase in the stabilizer subspace of \(>10^{10^4}\) if the newly discovered upper bound \(\xi_{14} \le 1458\) is tight.

Examining the results in Table~\ref{tab:Gausssumresults}, a deviation from the relationship \(2^{\alpha t} \leftrightarrow 3^{\alpha t}\) can be observed for the tensor upper bounds \(\chi_k^{t/k}\) and \(\xi_k^{t/k}\) for \(k>2\). A similar deviation was found for a related measure to the stabilizer rank, the \emph{approximate} stabilizer rank, of qutrits compared to qubits in another study~\cite{Huang18}. The simplest explanation for this is the conversion issue that can occur due to the exponential factor \(\alpha\) being a real number while \(\chi_k\) and \(\xi_k\) are constrained to be integers; the qubit \(\chi_k\) for \(k > 2\) that lead to better tensor upper bounds are no longer powers of \(2\) and so \(3^{\log_3(\chi_k) t}\) cannot be an integer \(\forall t \in \mathbb Z\), as required for the qutrit \(\chi_k^{t/k}\) or \(\xi_k^{t/k}\).

This deviation also coincides with the observed behavior that the optimal qubit stabilizer decompositions for \(t>2\) no longer consist of only orthogonal stabilizer states that are Clifford-separable and have equiprobable weights~\cite{Bravyi16_2,Kocia18_2}. This property was central to the proof that established the one-to-one correspondence between the states making up an optimal stabilizer decomposition and quadratic Gauss sums in~\cite{Kocia18_2}.

This raises the possibility that \(\xi_k \ne \chi_k\) for \(k>2\). Moreover, perhaps choosing stabilizer states as the \(\mathcal O(n^3)\)-cost basis in a state expansion for \(n\) qudits is inequivalent in the average case compared to quadratic Gauss sums as the \(\mathcal O(n^3)\)-cost basis. In this study, we have only considered their equivalence in terms of worst-case hardness. We leave the pursuit of these answers for future consideration.

\paragraph*{\bf{\emph{Conclusion}}} In this study we found that the cost of classical strong simulation of universal quantum circuits with qutrit \(T\) gate magic states using the WWM formalism, which produces a linear combination of terms that are cost-equivalent to stabilizer decompositions, exhibits novel reductions for \(t=1\), \(2\), \(3\) and \(6\) qutrits, in agreement with the qubit case. In addition, as this is an algebraic method that is significantly more tractable than numerical search for stabilizer rank by Monte Carlo methods, we are able to derive simulation cost bounds up to \(t=14\) qutrit magic states and find another improvement to the trivial tensor bound from the \(12\)-qutrit \(T\) gate magic state. Numerical implementation of this method may allow for increasing this search to even larger \(t\) values.

\noindent---
L.K. thanks the NRC Fellowship and the National Institute of Standards and Technology where part of this work was done. 
This material is based upon work supported by the U.S. Department of Energy, Office of Science, Office of Advanced Scientific Computing Research, under the Quantum Computing Application Teams program. 
Sandia National Laboratories is a multimission laboratory managed and operated by National Technology \& Engineering Solutions of Sandia, LLC, a wholly owned subsidiary of Honeywell International Inc., for the U.S. Department of Energy's National Nuclear Security Administration under contract {DE-NA0003525}. This paper describes objective technical results and analysis. Any subjective views or opinions that might be expressed in the paper do not necessarily represent the views of the U.S. Department of Energy or the United States Government. SAND2020-2655 O.

\bibliography{biblio}{}
\bibliographystyle{unsrt}

\onecolumngrid
\appendix

\section{Invariance of Eq.~\ref{eq:applicationtrace} under Linear Transformation of Final or Intermediate Variables}
\label{app:CNOTfreedom}

As described in~\cite{Kocia18_2}, the prime-\(d\) exponential sum, for arguments in \(\mathbb Z\), is invariant under some linear transformation \(\bsmc M\):
\begin{eqnarray}
  \label{eq:traceinvariance1}
  && \sum_{\substack{\bs y_q \in\\ (\mathbb Z / d \mathbb Z)^m}} \exp \frac{2 \pi i}{d} P(\bsmc M \bs y_q, \bs x_p, \bs x_q) \\
  &=& \sum_{\substack{\bsmc M^{-1} \bs y_q \in\\ (\mathbb Z / d \mathbb Z)^m}} \exp \frac{2 \pi i}{d} P(\bs y_q, \bs x_p, \bs x_q) \nonumber\\
  &=& \sum_{\substack{\bs y_q \in\\ (\mathbb Z / d \mathbb Z)^m}} \exp \frac{2 \pi i}{d} P(\bs y_q, \bs x_p, \bs x_q). \nonumber
\end{eqnarray}
This is because a linear transformation over the domain of a field merely permutes the order of the sum.

It further follows that, given a marginal trace over a single degree of freedom, a linear transformation \(\bsmc M\) merely changes the degree of freedom that is traced over:
\begin{eqnarray}
  \label{eq:traceinvariance2}
  && \sum_{\substack{\bs y_q \in\\ (\mathbb Z / d \mathbb Z)^m}} \sum_{\substack{\bs x \in\\ (\mathbb Z / d \mathbb Z)^{2m}\\ \bs x_{m+1} = 0}} e^{\frac{2 \pi i}{d} P(\bs y_q, \bsmc M \bs x)} \\
  &=& \sum_{\substack{\bs y_q \in\\ (\mathbb Z / d \mathbb Z)^m}} \sum_{\substack{\bsmc M^{-1} \bs x \in\\ (\mathbb Z / d \mathbb Z)^{2m} \\ (\bsmc M^{-1} \bs x)_{m+1} = 0}} e^{\frac{2 \pi i}{d} P(\bs y_q, \bs x)}. \nonumber\\
  &=& \sum_{\substack{\bs y_q \in\\ (\mathbb Z / d \mathbb Z)^m}} \sum_{\substack{\bs x \in\\ (\mathbb Z / d \mathbb Z)^{2m} }} e^{\frac{2 \pi i}{d} P(\bs y_q, (\bs x_p, \bs x_q))} \delta ((\bsmc M^{-1} \bs x)_{m+1}), \nonumber
\end{eqnarray}
where again, the final simplifcation results from the recognizing that the linear transformation only permutes the order of the sum.

These identities generalize to displaced linear transformations (affine transformations)~\cite{Kocia18_2}. The latter capture Clifford transformations, which form a symplectic subgroup.

Eq.~\ref{eq:applicationtrace} is of the form of Eq.~\ref{eq:traceinvariance2}. We investigate the minimal number of terms in its sum after any Clifford transformation \(\hat U_C\) in Eq.~\ref{eq:applicationtrace}. Therefore, due to Eq.~\ref{eq:traceinvariance1} and Eq.~\ref{eq:traceinvariance2}, it follows that we are free to additionally transform Eq.~\ref{eq:applicationtrace}'s \(\bs y_q\) and \(\bs x\) variables by a linear tranformation corresponding to Clifford transformations without affecting the worst-case analysis of the minimum number of sums.

Here, we find that products of the Clifford controlled-not gate are sufficient for our purposes.

The choice of a sequence of controlled-not transformations is not unique and only serves to make any reduction in the number of quadratic Gauss sums to be more easily recognized in an algebraic analysis. Generally, we choose a transformation so that half of the qudit degrees of freedom are ``cubic'' variables and the other half are ``quadratic'' variables---for fixed cubic variables, the remaining variables form a quadratic Gauss sum. This allows the ``cubic'' variables to function as indices that label the quadratic Gauss sums and determine their covariance and linear coefficients.

Further controlled-not transformations are made to reduce the number of quadratic Gauss sums by transforming the coefficients of some cubic indexing variables so that they only depend on other cubic indexing variables. This allows for repetitions or simplifications to become apparent without relying on quadratic Gauss sum identities. This allows only the indexing variables to be in the arguments of any additional Kronecker delta functions that reduce the number of quadratic Gauss sums that must be included in the full sum.

\section{Two Qutrit Magic State}
\label{app:twoqutritmagicstate}

We transform the initial and final variables of \(\rho_{T^{\otimes 2}}\) with \(C^2_{1,2}\) to obtain Eq.~\ref{eq:twoqutritpi8gate_2}.
The full form of \(\mathcal A_2\) in Eq.~\ref{eq:twoqutritpi8gate_2} is:
\begin{eqnarray}
  \mathcal A_2(y_{q_1}, \bs x)&=& \sum_{y_{q_2} \in \mathbb Z/ 3 \mathbb Z} \exp \left\{ \frac{2 \pi i}{3^2} \left[ 3 x_{q_1}^2 x_{q_2} + 6 x_{q_1} x_{q_2}^2 + 6 x_{q_1}^2 y_{q_1} + 6 x_{q_1} x_{q_2} y_{q_1} \right] \right\} \nonumber\\
    && \times \exp \left\{ \frac{2 \pi i}{3^2} \left[ 6 x_{q_2}^2 y_{q_1} + 6 x_{q_1} y_{q_1}^2 + 3 x_{q_2} y_{q_1}^2 + 3 x_{q_1}^2 y_{q_2} + 3 x_{q_1} x_{q_2} y_{q_2} \right] \right\} \nonumber\\
    && \times \exp \left\{ \frac{2 \pi i}{3^2} \left[ 6 x_{q_1} y_{q_1} y_{q_2} + 3 x_{q_2} y_{q_1} y_{q_2} + 6 y_{q_1}^2 y_{q_2} + 6 x_{q_1} y_{q_2}^2 + 3 y_{q_1} y_{q_2}^2 \right] \right\} \nonumber\\
    && \times \exp \left\{ \frac{2 \pi i}{3^2} \left[ x_{p_1} (6 y_{q_1} + 3 x_{q_1}) + x_{p_2} (6 y_{q_2} + 3 x_{q_2}) \right] \right\} \nonumber
\end{eqnarray}

\section{Three Qutrit Magic State}
\label{app:threequtritmagicstate}

  We act on the initial state with \(C_{1,2}^2\), \(C_{1,3}^2\), and \(C_{2,3}\) (the overall transformation we call \(C_3\)), which transforms \(y_{q_1} \rightarrow y_{q_1} - y_{q_2}\), \(y_{q_1} \rightarrow y_{q_1} - y_{q_3}\), and \(y_{q_2} \rightarrow y_{q_2} + y_{q_3}\), respectively. We also act on the final phase space variables with the same operators.

  This produces:
  \begin{eqnarray}
    \label{eq:threeweylpi8gatequtrit}
 \rho({\bsmc M}_{C_3} \bs x) &=& \sum_{\substack{y_{q_1}, y_{q_2} \\ \in \mathbb Z/ 3^2 \mathbb Z}} \exp \left[ \frac{2 \pi i}{9} \left( 7 y_{q_1}^3 + 8 x_{q_1}^3 \right) \right] \mathcal A_3 (y_{q_1}, y_{q_2}, y_{q_3}, \bs x) \nonumber
  \end{eqnarray}

\begin{eqnarray}
&& \mathcal A_3 (y_{q_1}, y_{q_2}, y_{q_3}, x_{p_1}, x_{p_2}, x_{p_3}, x_{q_1}, x_{q_2}, x_{q_3}) \\
 &=& \sum_{\substack{y_{q_3} \\ \in \mathbb Z/ 3^2 \mathbb Z}} \exp \Bigg\{ \frac{2 \pi i}{3} \Bigg[ - y_{q_1}^2 y_{q_2} + y_{q_1} y_{q_2}^2 - y_{q_2} x_{p_2} - y_{q_1}^2 x_{q_1} - y_{q_1} y_{q_2} x_{q_1} - y_{q_2}^2 x_{q_1} + x_{p_1} x_{q_1} \nonumber\\
&&  + y_{q_3}^2 (y_{q_1} - x_{q_1}) + y_{q_1}^2 x_{q_2} + y_{q_1} y_{q_2} x_{q_2} + x_{p_2} x_{q_2} + x_{q_1}^2 x_{q_2} - x_{q_1} x_{q_2}^2 + y_{q_2} x_{q_1} (x_{q_1} + x_{q_2}) \nonumber\\
&& - y_{q_1} (x_{p_1} + x_{q_1}^2 + x_{q_1} x_{q_2} + x_{q_2}^2) - y_{q_1}^2 x_{q_3} - y_{q_1} y_{q_2} x_{q_3} + y_{q_2}^2 x_{q_3} + x_{p_3} x_{q_3} - x_{q_1}^2 x_{q_3} \nonumber\\
&& - x_{q_1} x_{q_2} x_{q_3} + x_{q_2}^2 x_{q_3} - y_{q_2} (x_{q_1} + x_{q_2}) x_{q_3} + y_{q_1} (x_{q_1} - x_{q_2}) x_{q_3} - y_{q_1} x_{q_3}^2  - x_{q_1} x_{q_3}^2 - x_{q_3} + \Delta y_{q_3} \Bigg] \Bigg\}\nonumber
\end{eqnarray}
and
\begin{eqnarray}
\Delta &=& y_{q_1}^2 + y_{q_1} y_{q_2} - y_{q_2}^2 - x_{p_3} - x_{q_1}^2 - x_{q_1} x_{q_2} + x_{q_2}^2\\
&&  - y_{q_2} (x_{q_1} + x_{q_2}) + x_{q_1} x_{q_3} + y_{q_1} (x_{q_1} - x_{q_2} + x_{q_3} ) + 1. \nonumber
\end{eqnarray}

\section{Six Qutrit Magic State}
\label{app:sixqutritmagicstate}

After transforming the intermediate and final phase space variables by \(C_{1,2}^2\), \(C_{3,4}^2\), \(C_{5,6}^2\), \(C_{3,5}^2\), \(C_{1,3}^2\), \(C_{4,3}^2\), \(C_{6,5}^2\), \(C_{3,5}^2\), \(C_{6,3}^2\), \(C_{5,2}^2\), \(C_{2,4}^2\), \(C_{6,3}^2\), \(C_{4,5}\), \(C_{4,3}^2\), \(C_{3,2}^2\), \(C_{6,2}\), \(C_{5,3}^2\), \(C_{6,3}^2\), \(C_{2,1}\), \(C_{1,5}\), \(C_{3,5}\), \(C_{5,3}^2\), \(C_{3,5}\), and \(C_{1,3}\) (the overall transformation we call \(C_6\)), we find:

\begin{eqnarray}
  && \rho(\bsmc M_{C_6} (x_{p_1}, x_{p_2}, x_{p_3}, x_{p_4}, x_{p_5}, x_{p_6}, x_{q_1}, x_{q_2}, x_{q_3}, x_{q_4}, x_{q_5}, x_{q_6})) \nonumber\\
  &=& \sum_{\substack{y_{q_1}, y_{q_3}, y_{q_4}\\ \in \mathbb Z/ 3^2 \mathbb Z}} \exp \left[ \frac{2 \pi i}{9} \left( 4 y_{q_1}^3 + 2 x_{q_1}^3 \right) \right] \exp \left[ \frac{2 \pi i}{3} \Gamma_6(y_{q_1}, y_{q_3}, y_{q_4}, \bs x_p, \bs x_q) \right] \mathcal A_6(\bs y_{q}, \bs x_p, \bs x_q) 
\label{eq:sixweylpi8gatequtrit}
\end{eqnarray}
where
\begin{eqnarray}
  &&\Gamma_6(y_{q_1}, y_{q_3}, y_{q_4}, \bs x_p, \bs x_q) \\
  &\equiv& \Gamma_6(y_{q_1}, y_{q_3}, y_{q_4}, x_{p_1}, x_{p_2}, x_{p_3}, x_{p_4}, x_{p_5}, x_{p_6}, x_{q_1}, x_{q_2}, x_{q_3}, x_{q_4}, x_{q_5}, x_{q_6}) \nonumber\\
  &=& 2 y_{q_3}^2 y_{q_4} + y_{q_4}^3 + 2 y_{q_3} x_{p_3} + 2 y_{q_4} x_{p_4} + 2 y_{q_4}^2 x_{q_1} +  x_{p_1} x_{q_1} + 2 y_{q_4} x_{q_1}^2 + 2 y_{q_3} y_{q_4} x_{q_3} + x_{p_3} x_{q_3} +  y_{q_4} x_{q_3}^2 \nonumber\\
  && + (x_{p_4} + x_{q_1} (y_{q_4} + 2 x_{q_1}) + (y_{q_3} + x_{q_3})^2) x_{q_4} +  2 x_{q_1} x_{q_4}^2 + 2 x_{q_4}^3 + y_{q_1}^2 (y_{q_4} + 2 (x_{q_1} + x_{q_4}))  \nonumber\\
  && +  y_{q_1} (y_{q_4}^2 + 2 (x_{p_1} + x_{q_1}^2) + x_{q_1} x_{q_4} + 2 x_{q_4}^2 + y_{q_4} (x_{q_1} + x_{q_4})),\nonumber
\end{eqnarray}

\begin{eqnarray}
  &&\mathcal A_6(\bs y_{q}, \bs x_p, \bs x_q) \nonumber\\
  &=& \sum_{\substack{y_{q_2}, y_{q_5}, y_{q_6}\\ \in \mathbb Z/ 3^2 \mathbb Z}} \exp \Bigg\{ \frac{2 \pi i}{3} \Bigg[ x_{p_2} x_{q_2} + x_{q_2}^2 (2 y_{q_1} + 2 y_{q_4} + 2 x_{q_1} + 2 x_{q_4}) +  \Sigma_{y_{q_5}} y_{q_5}^2 \nonumber\\
  && \qquad \qquad \qquad \qquad \qquad + \Sigma_{y_{q_6}} y_{q_6}^2 + \Sigma_{y_{q_2}} y_{q_2}^2 + \Delta_{y_{q_2}} y_{q_2} + (y_{q_3} + 2 y_{q_4} + x_{q_3} + 2 x_{q_4}) x_{q_5}^2 + \Delta_{2q_5} y_{q_5} \\
  && \qquad \qquad \qquad \qquad \qquad + x_{p_5} x_{q_5} + x_{p_6} x_{q_6}+ 2 (y_{q_3} + y_{q_4} + x_{q_3} + x_{q_4}) x_{q_6}^2 + \Sigma_{y_{q_6}} y_{q_6} \Bigg] \Bigg\}, \nonumber
\end{eqnarray}
for
\begin{eqnarray}
  \Sigma_{y_{q_2}} &=& y_{q_1} + y_{q_4} + 2 (x_{q_1} + x_{q_4}),\\
  \Sigma_{y_{q_5}} &=& 2 y_{q_3} + y_{q_4} + x_{q_3} + 2 x_{q_4},\\
  \Sigma_{y_{q_6}} &=& y_{q_3} + y_{q_4} + 2 x_{q_3} + 2 x_{q_4},\\
  \Delta_{y_{q_2}} &=& 2 x_{p_2} + x_{q_2} (y_{q_1} + y_{q_4} + x_{q_1} + x_{q_4}),\\
  \Delta_{y_{q_5}} &=& 2 x_{p_5} + (2 y_{q_3} + y_{q_4} + 2 x_{q_3} + x_{q_4}) x_{q_5},
\end{eqnarray}
and
\begin{equation}
    \Delta_{y_{q_6}} = 2 x_{p_6} + (y_{q_3} + y_{q_4} + x_{q_3} + x_{q_4}) x_{q_6}.
\end{equation}

In Eq.~\ref{eq:sixweylpi8gatequtrit}, the intermediate variables \(y_{q_2}\), \(y_{q_5}\), and \(y_{q_6}\) are quadratic while \(y_{q_1}\), \(y_{q_3}\), and \(y_{q_4}\) are cubic. \(y_{q_1}\) is the only intermediate variable that lies in the full \(9\)-cycle and with respect to the intermediate variables it only has cross-terms with the cubic ones. Hence, \(y_{q_1}\) indexes the quadratic sums over the intermediate quadratic variables in terms of \(3\)-cocycles \(\{0,3,6\}\), \(\{1,4,7\}\), and \(\{2,5,8\}\). The three cubic variables each take three non-periodic values which leads to \(3^3 = 9 \times 3\) quadratic Gauss sums (that are \(3-\)dimensional). However, we can reduce this number by noticing some properties.

If the linear coefficient of \(y_{q_5}\) or \(y_{q_6}\) is non-zero anywhere, it is non-zero for at least three values of (\(y_{q_3}\), \(y_{q_4}\)) independent of \(y_{q_1}\). If the linear coefficient of \(y_{q_2}\) is non-zero anywhere, it is non-zero for at least three values of (\(y_{q_1}\), \(y_{q_4}\)) independent of \(y_{q_3}\). Otherwise, \(y_{q_1} = x_{q_1}\), \(y_{q_3} \ne x_{q_3}\) gives you a set of quadratic Gauss sums (indexed by \(y_{q_4}\) and \(y_{q_3}\)) that must add up to a real number because \(y_{q_3}\)'s quadratic coefficient is w.r.t. \(x_{q_3}\) and so is equal for \(y_{q_3}-x_{q_3} \ne 0\) and so only its linear coefficient differs (linearly) meaning that any imaginary parts must cancel out when running through all values of its quadratic coefficient: (\(y_{q_4}-x_{q_4}\)). Hence, \(y_{q_3} \ne x_{q_3}\) index two sets of quadratic Gauss sums indexed by \(y_{q_4}\) that each sum up to the same total. Thus, it is sufficient to sum up one set and multiply by two. This takes six quadratic Gauss sums and replaces them with three.

This can be summarized by the following equation:
\begin{eqnarray}
\label{eq:sixqutritsummary}
  && \rho_{T^{\otimes 6}}( \bsmc M_{C_6}\bs x) \\
  &=& \sum_{\substack{y_{q_1},  y_{q_3}, y_{q_4}\\ \in \mathbb Z/ 3^2 \mathbb Z}} e^{\frac{2 \pi i}{9} \left( 4 y_{q_1}^3 + 2 x_{q_1}^3 \right) } e^{ \frac{2 \pi i}{3} \Gamma_6(y_{q_1}, y_{q_3}, y_{q_4}, \bs x_p, \bs x_q) } \nonumber\\
  && \qquad \qquad \times \mathcal A_6(\bs y_{q}, \bs x_p, \bs x_q) \nonumber\\
  && \times \Bigg[ \delta( (\Sigma_{y_{q_2}} \land \Delta_{y_{q_2}}) \lor (\Sigma_{y_{q_5}} \land \Delta_{y_{q_5}}) \lor (\Sigma_{y_{q_6}} \land \Delta_{y_{q_6}})) \nonumber\\
  && \qquad + [\delta(y_{q_3} - x_{q_3}) + 2 \delta(y_{q_3} - x_{q_3} + 1)] \nonumber\\
  && \qquad \qquad \times \delta( \neg( \Sigma_{y_{q_2}} \lor \Sigma_{y_{q_5}} \lor \Sigma_{y_{q_6}})) \delta(y_{q_1}-x_{q_1})  \nonumber\\
  && \qquad + \delta( \neg( \Sigma_{y_{q_2}} \lor \Sigma_{y_{q_5}} \lor \Sigma_{y_{q_6}} )) \delta(\neg(y_{q_1}-x_{q_1})) \Bigg],\nonumber
\end{eqnarray}
where 
\begin{eqnarray}
  \delta(\alpha \lor \beta) &=& \delta(\alpha) \delta(\neg \beta) + \delta(\neg \alpha) \delta(\beta) \\
                            && + \delta(\alpha) \delta(\beta), \, \text{(logical inclusive disjunction)} \nonumber
\end{eqnarray}
and
\begin{equation}
  \delta(\alpha \land \beta) = \delta(\alpha) \delta(\beta), \, \text{(logical conjunction)}
\end{equation}
and the arguments for the delta functions are taken mod \(p\) odd-prime.

In Eq.~\ref{eq:sixqutritsummary}, the first term includes all cases where the quadratic coefficients of \(y_{q_2}\), \(y_{q_5}\), and \(y_{q_6}\) are zero along with their respective linear coefficients, as these produce plane waves that do not evaluate to zero. The second term includes all cases where all these quadratic coefficients are non-zero and \(y_{q_1} = x_{q_1}\) when the two sets of quadratic Gauss sums indexed by \(y_{q_3} \ne x_{q_3}\) sum up to the same value. The third term includes the remaining terms when the quadratic coefficients of \(y_{q_2}\), \(y_{q_5}\), and \(y_{q_6}\) are non-zero and \(y_{q_1} \ne x_{q_1}\). 

As before for three-qutrit \(T\) gate magic state, these three terms are disjoint--only one term is non-zero given \((\bs x_p, \bs x_q)\) and \(\bs y_{q}\). However, from the discussion earlier, there are fewer terms here than for the unfettered sum over \(\bs y_{q}\); the number of quadatic Gauss sums is reduced by three.

The same sort of analysis can be found for the corresponding final phase space variables.

\section{Twelve Qutrit Magic State}
\label{app:twelvequtritmagicstate}

After transforming the intermediate and final phase space variables by \(C_{12,11}^2\), \(C_{10,9}^2\), \(C_{8,7}^2\), \(C_{6,5}^2\), \(C_{4,3}^2\), \(C_{2,1}^2\), \(C_{10,12}^2\) \(C_{8,10}^2\), \(C_{6,8}^2\), \(C_{4,6}^2\), \(C_{2,4}^2\), \(C_{12,9}^2\), \(C_{9,12}\), \(C_{9,6}^2\), \(C_{6,9}\), \(C_{10,7}^2\), \(C_{7,10}\), \(C_{3,5}^2\), \(C_{7,5}^2\), \(C_{5,7}\), \(C_{3,1}^2\), \(C_{1,3}\), \(C_{5,3}^2\), \(C_{3,5}\), \(C_{8,5}^2\), \(C_{5,8}\), \(C_{9,7}\), \(C_{1,3}^2\), \(C_{5,3}\), \(C_{4,5}\), \(C_{6,3}\), \(C_{5,3}^2\), \(C_{3,5}\), \(C_{5,6}\), \(C_{6,1}^2\), \(C_{11,6}\), \(C_{6,1}^2\), \(C_{11,1}\), \(C_{11,2}\), \(C_{11,3}\), \(C_{11,4}^2\), \(C_{7,11}\), \(C_{7,5}\), \(C_{11,7}^2\), and \(C_{7,11}\), (the overall transformation we call \(C_{12}\)), we find:

\begin{eqnarray}
  && \rho(\bsmc M_{C_{12}} (x_{p_1}, \ldots, x_{p_{12}}, x_{q_1}, \ldots, x_{q_{12}})) \nonumber\\
  \label{eq:twelveweylpi8gatequtrit}
  &=& \sum_{\substack{y_{q_1}, \ldots, y_{q_6}\\ \in \mathbb Z/ 3^2 \mathbb Z}} \exp \left[ \frac{2 \pi i}{9} \left( 7 y_{q_2}^3 + 8 x_{q_2}^3 \right) \right] \exp \left[ \frac{2 \pi i}{3} \Gamma_{12}(y_{q_1}, \ldots , y_{q_6}, \bs x_p, \bs x_q) \right] \mathcal A_{12}(\bs y_{q}, \bs x_p, \bs x_q),
\end{eqnarray}
where
\begin{eqnarray}
  &&\Gamma_{12}(y_{q_1}, \ldots , y_{q_6}, \bs x_p, \bs x_q) =\\
&& 2 y_{q_3}^3 + 2 y_{q_3}^2 y_{q_4} + y_{q_3} y_{q_4}^2 + 2 y_{q_3}^2 y_{q_5} + 
 2 y_{q_3} y_{q_4} y_{q_5} + 2 y_{q_3} y_{q_5}^2 + 2 y_{q_3}^2 y_{q_6} + 2 y_{q_3} y_{q_4} y_{q_6} \nonumber\\
  && + y_{q_3} x_{p_{11}} + y_{q_5} x_{p_{11}} + y_{q_6} x_{p_{11}} + y_{q_5} x_{p_3} + y_{q_6} x_{p_3} + 
 y_{q_3} x_{p_4} + 2 y_{q_4} x_{p_4} + y_{q_5} x_{p_4} + y_{q_6} x_{p_4} + 2 y_{q_3} x_{p_5} \nonumber\\
  && + y_{q_5} x_{p_5} + 2 y_{q_6} x_{p_5} + 2 y_{q_6} x_{p_6} + 2 y_{q_3} y_{q_4} x_{q_1} + 
 2 y_{q_3} y_{q_5} x_{q_1} + 2 y_{q_5} y_{q_6} x_{q_1} + x_{p_1} x_{q_1} + 2 x_{p_{11}} x_{q_1} \nonumber\\
  && + 2 x_{p_3} x_{q_1} + 2 x_{p_4} x_{q_1} + x_{p_5} x_{q_1} + y_{q_3}^2 x_{q_2} + y_{q_3} y_{q_4} x_{q_2} + 
 2 y_{q_4}^2 x_{q_2} + 2 y_{q_3} y_{q_5} x_{q_2} + 2 y_{q_4} y_{q_6} x_{q_2} + 2 y_{q_6}^2 x_{q_2} \nonumber\\
  && + x_{p_2} x_{q_2} + y_{q_3} x_{q_1} x_{q_2} + y_{q_3} x_{q_2}^2 + y_{q_4} x_{q_2}^2 + 2 y_{q_6} x_{q_2}^2 + 
 2 y_{q_3} y_{q_4} x_{q_3} + 2 y_{q_4}^2 x_{q_3} + 2 y_{q_3} y_{q_5} x_{q_3} + y_{q_4} y_{q_5} x_{q_3} \nonumber\\
  && + y_{q_5}^2 x_{q_3} + 2 y_{q_3} y_{q_6} x_{q_3} + y_{q_4} y_{q_6} x_{q_3} + 2 x_{p_{11}} x_{q_3} + 
 2 x_{p_4} x_{q_3} + x_{p_5} x_{q_3} + 2 y_{q_4} x_{q_1} x_{q_3} + 2 y_{q_5} x_{q_1} x_{q_3} \nonumber\\
  && + 2 y_{q_3} x_{q_2} x_{q_3} + y_{q_4} x_{q_2} x_{q_3} + 2 y_{q_5} x_{q_2} x_{q_3} + x_{q_1} x_{q_2} x_{q_3} + 
 x_{q_2}^2 x_{q_3} + y_{q_4} x_{q_3}^2 + y_{q_5} x_{q_3}^2 + y_{q_6} x_{q_3}^2 + x_{q_2} x_{q_3}^2 + x_{q_3}^3 \nonumber\\
  && + y_{q_3}^2 x_{q_4} + y_{q_3} y_{q_4} x_{q_4} + y_{q_3} y_{q_5} x_{q_4} + y_{q_3} y_{q_6} x_{q_4} + x_{p_4} x_{q_4} + 2 y_{q_3} x_{q_1} x_{q_4} + y_{q_3} x_{q_2} x_{q_4} + y_{q_4} x_{q_2} x_{q_4} + 2 y_{q_6} x_{q_2} x_{q_4} \nonumber\\
  && + x_{q_2}^2 x_{q_4} + 2 y_{q_3} x_{q_3} x_{q_4} + y_{q_4} x_{q_3} x_{q_4} + y_{q_5} x_{q_3} x_{q_4} + y_{q_6} x_{q_3} x_{q_4} + 2 x_{q_1} x_{q_3} x_{q_4} + 
 x_{q_2} x_{q_3} x_{q_4} + x_{q_3}^2 x_{q_4} + 2 y_{q_3} x_{q_4}^2 + 2 x_{q_2} x_{q_4}^2 \nonumber\\
  &&+ 2 x_{q_3} x_{q_4}^2 + y_{q_3}^2 x_{q_5} + 2 y_{q_3} y_{q_4} x_{q_5} + y_{q_4}^2 x_{q_5} + 2 y_{q_4} y_{q_5} x_{q_5} + 
 2 y_{q_5}^2 x_{q_5} + y_{q_3} y_{q_6} x_{q_5} + 2 y_{q_4} y_{q_6} x_{q_5} + 2 x_{p_3} x_{q_5} \nonumber\\
  && + x_{p_5} x_{q_5} + 2 y_{q_3} x_{q_1} x_{q_5} + y_{q_4} x_{q_1} x_{q_5} + y_{q_5} x_{q_1} x_{q_5} + 
 2 y_{q_6} x_{q_1} x_{q_5} + 2 y_{q_4} x_{q_2} x_{q_5} + y_{q_5} x_{q_2} x_{q_5} + 2 x_{q_1} x_{q_2} x_{q_5} \nonumber\\
  && + 2 x_{q_2}^2 x_{q_5} + 2 y_{q_3} x_{q_3} x_{q_5} + 2 y_{q_4} x_{q_3} x_{q_5} + y_{q_6} x_{q_3} x_{q_5} + 
 2 x_{q_1} x_{q_3} x_{q_5} + x_{q_3}^2 x_{q_5} + 2 y_{q_3} x_{q_4} x_{q_5} + 2 y_{q_4} x_{q_4} x_{q_5} \nonumber\\
  && + 2 y_{q_5} x_{q_4} x_{q_5} + 2 y_{q_6} x_{q_4} x_{q_5} + x_{q_1} x_{q_4} x_{q_5} + 2 x_{q_2} x_{q_4} x_{q_5} + 
 2 x_{q_3} x_{q_4} x_{q_5} + x_{q_4}^2 x_{q_5} + 2 y_{q_3} x_{q_5}^2 + 2 y_{q_5} x_{q_5}^2 \nonumber\\
  && + y_{q_6} x_{q_5}^2 + x_{q_1} x_{q_5}^2 + 2 x_{q_2} x_{q_5}^2 + 2 x_{q_3} x_{q_5}^2 + 
 2 x_{q_5}^3 + (y_{q_3}^2 + 2 x_{p_{11}} + 2 x_{p_3} + 2 x_{p_4} + x_{p_5} + x_{p_6} + 2 y_{q_5} x_{q_1} \nonumber\\
  && + x_{q_2} (y_{q_6} + 2 x_{q_2}) + x_{q_3}^2 + (2 x_{q_2} + x_{q_3}) x_{q_4} + y_{q_3} (y_{q_4} + 2 x_{q_3} + x_{q_4} + x_{q_5}) + 
    x_{q_5} (2 x_{q_1} + x_{q_3} + 2 x_{q_4} + x_{q_5}) \nonumber\\
  && + y_{q_4} (2 x_{q_2} + x_{q_3} + 2 x_{q_5})) x_{q_6} + 2 x_{q_2} x_{q_6}^2 + y_{q_1} (y_{q_5} y_{q_6}
     + 2 x_{p_1} + x_{p_{11}} + x_{p_3} + x_{p_4} + 2 x_{p_5} + 2 y_{q_4} x_{q_3} + 2 y_{q_5} x_{q_3} + x_{q_2} x_{q_3}\nonumber\\
  && + 2 x_{q_3} x_{q_4} + y_{q_4} x_{q_5} + y_{q_5} x_{q_5} + 2 y_{q_6} x_{q_5} 
   + 2 x_{q_2} x_{q_5} + 2 x_{q_3} x_{q_5} + x_{q_4} x_{q_5} + x_{q_5}^2 + y_{q_2} (2 y_{q_3} + x_{q_3} + 2 x_{q_5}) \nonumber\\
  && + y_{q_3} (y_{q_4} + y_{q_5} + x_{q_2} + 2 (x_{q_4} + x_{q_5})) + 2 (y_{q_5} + x_{q_5}) x_{q_6}) 
  + y_{q_2}^2 (2 y_{q_3} + 2 y_{q_4} + y_{q_6} + 2 x_{q_2} + x_{q_3} + x_{q_4} + 2 (x_{q_5} + x_{q_6})) \nonumber\\
  && + y_{q_2} (2 y_{q_3}^2 + y_{q_4}^2 + y_{q_6}^2 + 2 x_{p_2} + y_{q_6} x_{q_2} + 2 x_{q_2}^2 
  + 2 y_{q_5} x_{q_3} + x_{q_1} x_{q_3} + 2 x_{q_2} x_{q_3} + x_{q_3}^2 + 2 y_{q_6} x_{q_4} + 2 x_{q_2} x_{q_4} \nonumber\\
  && + x_{q_3} x_{q_4} + 2 x_{q_4}^2 + y_{q_3} (2 y_{q_4} + y_{q_5} + x_{q_1} + 2 (x_{q_2} + x_{q_3}) + x_{q_4}) + y_{q_5} x_{q_5} 
  + 2 x_{q_1} x_{q_5} + x_{q_2} x_{q_5} + 2 x_{q_4} x_{q_5} + 2 x_{q_5}^2 + \nonumber\\
  && (y_{q_6} + x_{q_2} + 2 x_{q_4}) x_{q_6} + 2 x_{q_6}^2 + y_{q_4} (y_{q_6} + 2 x_{q_2} + x_{q_3} + x_{q_4} + 2 (x_{q_5} + x_{q_6}))) \nonumber
\end{eqnarray}
\begin{eqnarray}
  &&\mathcal A_{12}(\bs y_{q}, \bs x_p, \bs x_q) \\
  &=& \sum_{\substack{y_{q_7}, \ldots, y_{q_{12}}\\ \in \mathbb Z/ 3^2 \mathbb Z}} \exp \Bigg\{ \frac{2 \pi i}{3} \Bigg[ \Sigma_{y_{q_{12}}} y_{q_{12}}^2 + x_{q_{12}}^2 (2 y_{q_3} + 2 y_{q_5} + 2 x_{q_3})  + \Sigma_{y_{q_{10}}} y_{q_{10}}^2 + x_{q_{10}}^2 (2 y_{q_3} + 2 x_{q_3} + x_{q_5}) + \Sigma_{y_{q_8}} y_{q_8}^2\nonumber\\
  && \qquad \qquad \qquad \qquad \qquad + \Sigma_{y_{q_9}} y_{q_9}^2 + \Sigma_{y_{q_7}}(y_{q_6}) y_{q_7}^2 + \Sigma_{y_{q_{11}}}(y_{q_6}) y_{q_{11}}^2 + x_{q_{11}}^2 (2 y_{q_3} + 2 y_{q_5} + 2 y_{q_6} + 2 x_{q_3} + 2 x_{q_6}) \nonumber\\
  && \qquad \qquad \qquad \qquad \qquad + x_{q_{12}} (y_{q_1} y_{q_3} + y_{q_1} y_{q_5} + y_{q_3} y_{q_6} + y_{q_5} y_{q_6} + x_{p_{12}} + 
    y_{q_3} x_{q_1} + y_{q_5} x_{q_1} + y_{q_1} x_{q_3}\nonumber\\
  && \qquad \qquad \qquad \qquad \qquad + y_{q_6} x_{q_3} + x_{q_1} x_{q_3} + y_{q_3} x_{q_6} + 
    y_{q_5} x_{q_6} + x_{q_3} x_{q_6}) + \Delta_{y_{q_{12}}} y_{q_{12}} \nonumber\\
  && \qquad \qquad \qquad \qquad \qquad + x_{q_{11}} (2 y_{q_1} y_{q_3} + 2 y_{q_1} y_{q_5} + y_{q_3} y_{q_5} + y_{q_5}^2 + 2 y_{q_1} y_{q_6} + 
    y_{q_5} y_{q_6} + x_{p_{11}} + 2 y_{q_3} x_{q_1} \nonumber\\
  && \qquad \qquad \qquad \qquad \qquad + 2 y_{q_5} x_{q_1} + 2 y_{q_6} x_{q_1} + 
    2 y_{q_1} x_{q_3} + y_{q_5} x_{q_3} + 2 x_{q_1} x_{q_3} + y_{q_3} x_{q_5} + y_{q_5} x_{q_5} + y_{q_6} x_{q_5} + x_{q_3} x_{q_5} \nonumber\\
  && \qquad \qquad \qquad \qquad \qquad + 2 y_{q_1} x_{q_6} + y_{q_5} x_{q_6} + 2 x_{q_1} x_{q_6} + 
    x_{q_5} x_{q_6}) + 
 x_{q_{10}} (y_{q_1} y_{q_3} + 2 y_{q_3}^2 + y_{q_3} y_{q_5} + y_{q_3} y_{q_6} \nonumber\\
  && \qquad \qquad \qquad \qquad \qquad + x_{p_{10}} + y_{q_3} x_{q_1} + y_{q_1} x_{q_3} + y_{q_3} x_{q_3} + y_{q_5} x_{q_3} + y_{q_6} x_{q_3} + x_{q_1} x_{q_3} + 
    2 x_{q_3}^2 + 2 y_{q_1} x_{q_5} + 2 y_{q_5} x_{q_5}\nonumber\\
  && \qquad \qquad \qquad \qquad \qquad + 2 y_{q_6} x_{q_5} + 2 x_{q_1} x_{q_5} + x_{q_5}^2 + y_{q_3} x_{q_6} + x_{q_3} x_{q_6} + 2 x_{q_5} x_{q_6}) + \Delta_{y_{q_{10}}} y_{q_{10}}  + \Delta_{y_{q_{11}}} y_{q_{11}} \nonumber\\
  && \qquad \qquad \qquad \qquad \qquad+ (2 y_{q_1}^2 + y_{q_1} y_{q_2} + 2 y_{q_2} y_{q_3} + y_{q_3}^2 + 2 y_{q_1} y_{q_4} + y_{q_3} y_{q_4} + 2 y_{q_1} y_{q_5} + 2 y_{q_2} y_{q_5} \nonumber\\
  && \qquad \qquad \qquad \qquad \qquad + y_{q_4} y_{q_5} + 2 y_{q_5}^2 + 2 y_{q_1} y_{q_6} + y_{q_2} y_{q_6} + 2 y_{q_3} y_{q_6} + 2 y_{q_4} y_{q_6} + y_{q_5} y_{q_6} + x_{p_7} + y_{q_1} x_{q_1} \nonumber\\
  && \qquad \qquad \qquad \qquad \qquad + y_{q_2} x_{q_1} + 2 y_{q_4} x_{q_1} + 2 y_{q_5} x_{q_1} + 2 y_{q_6} x_{q_1} + 2 x_{q_1}^2 + y_{q_1} x_{q_2} + 2 y_{q_3} x_{q_2} + 2 y_{q_5} x_{q_2} + y_{q_6} x_{q_2} \nonumber\\
  && \qquad \qquad \qquad \qquad \qquad + x_{q_1} x_{q_2} + 2 y_{q_2} x_{q_3} + 2 y_{q_3} x_{q_3} + y_{q_4} x_{q_3} + 2 y_{q_6} x_{q_3} + 2 x_{q_2} x_{q_3} + x_{q_3}^2 + 2 y_{q_1} x_{q_4} + y_{q_3} x_{q_4} \nonumber\\
  && \qquad \qquad \qquad \qquad \qquad + y_{q_5} x_{q_4} + 2 y_{q_6} x_{q_4} + 2 x_{q_1} x_{q_4} + x_{q_3} x_{q_4} + 2 y_{q_1} x_{q_5} + y_{q_3} x_{q_5} + y_{q_5} x_{q_5} + 2 y_{q_6} x_{q_5} + 2 x_{q_1} x_{q_5} \nonumber\\
  && \qquad \qquad \qquad \qquad \qquad + x_{q_3} x_{q_5} + 2 y_{q_1} x_{q_6} + y_{q_2} x_{q_6} + 2 y_{q_3} x_{q_6} + 2 y_{q_4} x_{q_6} + y_{q_5} x_{q_6} + 2 x_{q_1} x_{q_6} + x_{q_2} x_{q_6} + 2 x_{q_3} x_{q_6} \nonumber\\
  && \qquad \qquad \qquad \qquad \qquad + 2 x_{q_4} x_{q_6} + 2 x_{q_5} x_{q_6}) x_{q_7} + (y_{q_1} + 2 y_{q_3} + 2 y_{q_5} + y_{q_6} + x_{q_1} + 2 x_{q_3} + x_{q_6}) x_{q_7}^2 + \Delta_{y_{q_7}} y_{q_7} \nonumber\\
  && \qquad \qquad \qquad \qquad \qquad + (2 y_{q_1} y_{q_2} + y_{q_2}^2 + 2 y_{q_1} y_{q_3} + 2 y_{q_2} y_{q_3} + y_{q_3}^2 + y_{q_1} y_{q_4} + y_{q_2} y_{q_4} + y_{q_3} y_{q_4} \nonumber\\
  && \qquad \qquad \qquad \qquad \qquad + y_{q_4}^2 + y_{q_2} y_{q_5} + y_{q_3} y_{q_5} + 2 y_{q_4} y_{q_5} + y_{q_2} y_{q_6} + y_{q_3} y_{q_6} + 2 y_{q_4} y_{q_6} + x_{p_8} + 2 y_{q_2} x_{q_1}  \nonumber\\
  && \qquad \qquad \qquad \qquad \qquad + 2 y_{q_3} x_{q_1} + y_{q_4} x_{q_1} + 2 y_{q_1} x_{q_2} + 2 y_{q_2} x_{q_2} + 2 y_{q_3} x_{q_2} + y_{q_4} x_{q_2} + y_{q_5} x_{q_2} + y_{q_6} x_{q_2} + 2 x_{q_1} x_{q_2} \nonumber\\
  && \qquad \qquad \qquad \qquad \qquad + x_{q_2}^2 + 2 y_{q_1} x_{q_3} + 2 y_{q_2} x_{q_3} + 2 y_{q_3} x_{q_3} + y_{q_4} x_{q_3} + y_{q_5} x_{q_3} + y_{q_6} x_{q_3} + 2 x_{q_1} x_{q_3} + 2 x_{q_2} x_{q_3} \nonumber\\
  && \qquad \qquad \qquad \qquad \qquad + x_{q_3}^2 + y_{q_1} x_{q_4} + y_{q_2} x_{q_4} + y_{q_3} x_{q_4} + 2 y_{q_4} x_{q_4} + 2 y_{q_5} x_{q_4} + 2 y_{q_6} x_{q_4} + x_{q_1} x_{q_4} + x_{q_2} x_{q_4} \nonumber\\
  && \qquad \qquad \qquad \qquad \qquad + x_{q_3} x_{q_4} + x_{q_4}^2 + y_{q_1} x_{q_5} + 2 y_{q_2} x_{q_5} + 2 y_{q_3} x_{q_5} + y_{q_4} x_{q_5} + 2 y_{q_5} x_{q_5} + 2 y_{q_6} x_{q_5} + x_{q_1} x_{q_5} \nonumber\\
  && \qquad \qquad \qquad \qquad \qquad + 2 x_{q_2} x_{q_5} + 2 x_{q_3} x_{q_5} + x_{q_4} x_{q_5} + y_{q_2} x_{q_6} + y_{q_3} x_{q_6} + 2 y_{q_4} x_{q_6} + x_{q_2} x_{q_6} + x_{q_3} x_{q_6} + 2 x_{q_4} x_{q_6} \nonumber\\
  && \qquad \qquad \qquad \qquad \qquad + 2 x_{q_5} x_{q_6}) x_{q_8} + (2 y_{q_2} + 2 y_{q_3} + y_{q_4} + 2 x_{q_2} + 2 x_{q_3} + x_{q_4} + x_{q_5}) x_{q_8}^2 + \Delta_{y_{q_8}} y_{q_8} + (2 y_{q_1}^2 \nonumber\\
  && \qquad \qquad \qquad \qquad \qquad + 2 y_{q_1} y_{q_3} + y_{q_3} y_{q_4} + y_{q_4}^2 + y_{q_1} y_{q_6} + y_{q_3} y_{q_6} + y_{q_4} y_{q_6} + x_{p_9} + y_{q_1} x_{q_1} + 2 y_{q_3} x_{q_1} \nonumber\\
  && \qquad \qquad \qquad \qquad \qquad + y_{q_6} x_{q_1} + 2 x_{q_1}^2 + 2 y_{q_1} x_{q_3} + y_{q_4} x_{q_3} + y_{q_6} x_{q_3} + 2 x_{q_1} x_{q_3} + y_{q_3} x_{q_4} + 2 y_{q_4} x_{q_4} + y_{q_6} x_{q_4} \nonumber\\
  && \qquad \qquad \qquad \qquad \qquad + x_{q_3} x_{q_4} + x_{q_4}^2 + y_{q_1} x_{q_5} + 2 y_{q_4} x_{q_5} + 2 y_{q_6} x_{q_5} + x_{q_1} x_{q_5} + 2 x_{q_4} x_{q_5} + y_{q_1} x_{q_6} + y_{q_3} x_{q_6} \nonumber\\
  && \qquad \qquad \qquad \qquad \qquad + y_{q_4} x_{q_6} + x_{q_1} x_{q_6} + x_{q_3} x_{q_6} + x_{q_4} x_{q_6} + 2 x_{q_5} x_{q_6}) x_{q_9} + (2 y_{q_1} + 2 y_{q_3} + 2 y_{q_4} + 2 x_{q_1} + 2 x_{q_3} \nonumber\\
  && \qquad \qquad \qquad \qquad \qquad + 2 x_{q_4} + x_{q_5}) x_{q_9}^2 + \Delta_{y_{q_9}} y_{q_9} \Bigg] \Bigg\}. \nonumber
\end{eqnarray}
where
\begin{eqnarray}
  \Sigma_{y_{q_7}}(y_{q_6}) &=& (2 y_{q_1} + y_{q_3} + y_{q_5} + 2 y_{q_6} + x_{q_1} + 2 x_{q_3} + x_{q_6})\\
  \Sigma_{y_{q_8}} &=& (y_{q_2} + y_{q_3} + 2 y_{q_4} + 2 x_{q_2} + 2 x_{q_3} + x_{q_4} + x_{q_5})\\
  \Sigma_{y_{q_9}} &=& (y_{q_1} + y_{q_3} + y_{q_4} + 2 x_{q_1} + 2 x_{q_3} + 2 x_{q_4} + x_{q_5})\\
  \Sigma_{y_{q_{10}}} &=& (y_{q_3} + 2 x_{q_3} + x_{q_5})\\
  \Sigma_{y_{q_{11}}}(y_{q_6}) &=& (y_{q_3} + y_{q_5} + y_{q_6} + 2 x_{q_3} + 2 x_{q_6})\\
  \Sigma_{y_{q_{12}}} &=& (y_{q_3} + y_{q_5} + 2 x_{q_3})\\
  \Delta_{y_{q_7}} &=& (y_{q_1}^2 + 2 y_{q_1} y_{q_2} + y_{q_2} y_{q_3} + 2 y_{q_3}^2 + y_{q_1} y_{q_4} + 
    2 y_{q_3} y_{q_4} + y_{q_1} y_{q_5} + y_{q_2} y_{q_5} + 2 y_{q_4} y_{q_5} + y_{q_5}^2 \\
  && + y_{q_1} y_{q_6} + 2 y_{q_2} y_{q_6} + y_{q_3} y_{q_6} + y_{q_4} y_{q_6} + 2 y_{q_5} y_{q_6} + 
    2 x_{p_7} + y_{q_1} x_{q_1} + y_{q_2} x_{q_1} + 2 y_{q_4} x_{q_1} + 2 y_{q_5} x_{q_1} \nonumber\\
  && + 2 y_{q_6} x_{q_1} + 2 x_{q_1}^2 + y_{q_1} x_{q_2} + 2 y_{q_3} x_{q_2} + 2 y_{q_5} x_{q_2} + 
    y_{q_6} x_{q_2} + x_{q_1} x_{q_2} + 2 y_{q_2} x_{q_3} + 2 y_{q_3} x_{q_3} + y_{q_4} x_{q_3} \nonumber\\
  && + 2 y_{q_6} x_{q_3} + 2 x_{q_2} x_{q_3} + x_{q_3}^2 + 2 y_{q_1} x_{q_4} + y_{q_3} x_{q_4} + 
    y_{q_5} x_{q_4} + 2 y_{q_6} x_{q_4} + 2 x_{q_1} x_{q_4} + x_{q_3} x_{q_4} + 2 y_{q_1} x_{q_5} \nonumber\\
  && + y_{q_3} x_{q_5} + y_{q_5} x_{q_5} + 2 y_{q_6} x_{q_5} + 2 x_{q_1} x_{q_5} + x_{q_3} x_{q_5} + 
    2 y_{q_1} x_{q_6} + y_{q_2} x_{q_6} + 2 y_{q_3} x_{q_6} + 2 y_{q_4} x_{q_6} + y_{q_5} x_{q_6} \nonumber\\
  && + 2 x_{q_1} x_{q_6} + x_{q_2} x_{q_6} + 2 x_{q_3} x_{q_6} + 2 x_{q_4} x_{q_6} + 
    2 x_{q_5} x_{q_6} + (2 y_{q_1} + y_{q_3} + y_{q_5} + 2 y_{q_6} + 2 x_{q_1} + x_{q_3} + 2 x_{q_6}) x_{q_7}) \nonumber\\
  \Delta_{y_{q_8}} &=& (y_{q_1} y_{q_2} + 2 y_{q_2}^2 + y_{q_1} y_{q_3} + y_{q_2} y_{q_3} + 2 y_{q_3}^2 + 
    2 y_{q_1} y_{q_4} + 2 y_{q_2} y_{q_4} + 2 y_{q_3} y_{q_4} + 2 y_{q_4}^2 \\
  && + 2 y_{q_2} y_{q_5} + 2 y_{q_3} y_{q_5} + y_{q_4} y_{q_5} + 2 y_{q_2} y_{q_6} + 
    2 y_{q_3} y_{q_6} + y_{q_4} y_{q_6} + 2 x_{p_8} + 2 y_{q_2} x_{q_1} + 2 y_{q_3} x_{q_1} \nonumber\\
  && + y_{q_4} x_{q_1} + 2 y_{q_1} x_{q_2} + 2 y_{q_2} x_{q_2} + 2 y_{q_3} x_{q_2} + y_{q_4} x_{q_2} + 
    y_{q_5} x_{q_2} + y_{q_6} x_{q_2} + 2 x_{q_1} x_{q_2} + x_{q_2}^2 + 2 y_{q_1} x_{q_3} \nonumber\\
  && + 2 y_{q_2} x_{q_3} + 2 y_{q_3} x_{q_3} + y_{q_4} x_{q_3} + y_{q_5} x_{q_3} + y_{q_6} x_{q_3} + 
    2 x_{q_1} x_{q_3} + 2 x_{q_2} x_{q_3} + x_{q_3}^2 + y_{q_1} x_{q_4} + y_{q_2} x_{q_4} + y_{q_3} x_{q_4} \nonumber\\
  && + 2 y_{q_4} x_{q_4} + 2 y_{q_5} x_{q_4} + 2 y_{q_6} x_{q_4} + x_{q_1} x_{q_4} + x_{q_2} x_{q_4} + 
    x_{q_3} x_{q_4} + x_{q_4}^2 + y_{q_1} x_{q_5} + 2 y_{q_2} x_{q_5} + 2 y_{q_3} x_{q_5} + y_{q_4} x_{q_5} \nonumber\\
  && + 2 y_{q_5} x_{q_5} + 2 y_{q_6} x_{q_5} + x_{q_1} x_{q_5} + 2 x_{q_2} x_{q_5} + 2 x_{q_3} x_{q_5} + 
    x_{q_4} x_{q_5} + y_{q_2} x_{q_6} + y_{q_3} x_{q_6} + 2 y_{q_4} x_{q_6} + x_{q_2} x_{q_6} + x_{q_3} x_{q_6} \nonumber\\
  && + 2 x_{q_4} x_{q_6} + 2 x_{q_5} x_{q_6} + (y_{q_2} + y_{q_3} + 2 y_{q_4} + x_{q_2} + x_{q_3} + 2 x_{q_4} + 
       2 x_{q_5}) x_{q_8}) \nonumber\\
  \Delta_{y_{q_9}} &=& (y_{q_1}^2 + y_{q_1} y_{q_3} + 2 y_{q_3} y_{q_4} + 2 y_{q_4}^2 + 2 y_{q_1} y_{q_6} + 
    2 y_{q_3} y_{q_6} + 2 y_{q_4} y_{q_6} + 2 x_{p_9} + y_{q_1} x_{q_1} + 2 y_{q_3} x_{q_1} \\
  && + y_{q_6} x_{q_1} + 2 x_{q_1}^2 + 2 y_{q_1} x_{q_3} + y_{q_4} x_{q_3} + y_{q_6} x_{q_3} + 
    2 x_{q_1} x_{q_3} + y_{q_3} x_{q_4} + 2 y_{q_4} x_{q_4} + y_{q_6} x_{q_4} + x_{q_3} x_{q_4} + x_{q_4}^2 \nonumber\\
  && + y_{q_1} x_{q_5} + 2 y_{q_4} x_{q_5} + 2 y_{q_6} x_{q_5} + x_{q_1} x_{q_5} + 2 x_{q_4} x_{q_5} + 
    y_{q_1} x_{q_6} + y_{q_3} x_{q_6} + y_{q_4} x_{q_6} + x_{q_1} x_{q_6} + x_{q_3} x_{q_6} + x_{q_4} x_{q_6} \nonumber\\
  && + 2 x_{q_5} x_{q_6} + (y_{q_1} + y_{q_3} + y_{q_4} + x_{q_1} + x_{q_3} + x_{q_4} + 2 x_{q_5}) x_{q_9}) \nonumber\\
  \Delta_{y_{q_{10}}} &=& (2 y_{q_1} y_{q_3} + y_{q_3}^2 + 2 y_{q_3} y_{q_5} + 2 y_{q_3} y_{q_6} + 2 x_{p_{10}} + 
    y_{q_3} x_{q_1} + y_{q_1} x_{q_3} + y_{q_3} x_{q_3} + y_{q_5} x_{q_3} + y_{q_6} x_{q_3} \\
  && + x_{q_1} x_{q_3} + 2 x_{q_3}^2 + 2 y_{q_1} x_{q_5} + 2 y_{q_5} x_{q_5} + 2 y_{q_6} x_{q_5} + 2 x_{q_1} x_{q_5} + 
    x_{q_5}^2 + x_{q_{10}} (y_{q_3} + x_{q_3} + 2 x_{q_5}) + y_{q_3} x_{q_6} \nonumber\\
                && + x_{q_3} x_{q_6} + 2 x_{q_5} x_{q_6}) \nonumber\\
  \Delta_{y_{q_{11}}} &=& (y_{q_1} y_{q_3} + y_{q_1} y_{q_5} + 2 y_{q_3} y_{q_5} + 2 y_{q_5}^2 + y_{q_1} y_{q_6} + 
    2 y_{q_5} y_{q_6} + 2 x_{p_{11}} + 2 y_{q_3} x_{q_1} + 2 y_{q_5} x_{q_1} + 2 y_{q_6} x_{q_1} \\
  && + 2 y_{q_1} x_{q_3} + y_{q_5} x_{q_3} + 2 x_{q_1} x_{q_3} + y_{q_3} x_{q_5} + y_{q_5} x_{q_5} + 
    y_{q_6} x_{q_5} + x_{q_3} x_{q_5} + 2 y_{q_1} x_{q_6} + y_{q_5} x_{q_6} + 2 x_{q_1} x_{q_6} + x_{q_5} x_{q_6} \nonumber\\
  && + x_{q_{11}} (y_{q_3} + y_{q_5} + y_{q_6} + x_{q_3} + x_{q_6})) \nonumber
\end{eqnarray}
and
\begin{eqnarray}
  \Delta_{y_{q_{12}}} &=& (2 y_{q_1} y_{q_3} + 2 y_{q_1} y_{q_5} + 2 y_{q_3} y_{q_6} + 2 y_{q_5} y_{q_6} + 
    2 x_{p_{12}} + y_{q_3} x_{q_1} + y_{q_5} x_{q_1} + y_{q_1} x_{q_3} + y_{q_6} x_{q_3} + x_{q_1} x_{q_3} \nonumber\\
  && + x_{q_{12}} (y_{q_3} + y_{q_5} + x_{q_3}) + y_{q_3} x_{q_6} + y_{q_5} x_{q_6} + x_{q_3} x_{q_6}).
\end{eqnarray}

In Eq.~\ref{eq:twelveweylpi8gatequtrit}, the intermediate variables \(y_{q_7}\), \(\ldots\), \(y_{q_{12}}\) are quadratic while \(y_{q_1}\), \(\ldots\), \(y_{q_6}\) are cubic. \(y_{q_2}\) is the only intermediate coordinate that lies in the full \(9\)-cycle and with respect to the intermediate variables it only has cross-terms with the cubic ones. Hence, \(y_{q_2}\) indexes the quadratic sums over the intermediate quadratic variables in terms of \(3\)-cocycles \(\{0,3,6\}\), \(\{1,4,7\}\), and \(\{2,5,8\}\). The three cubic variables each take three non-periodic values which leads to \(3^6\) quadratic Gauss sums (that are \(3^6-\)dimensional). However, we can reduce this number by noticing some properties.

Given \(y_{q_1}\), \(y_{q_2}\), \(y_{q_3}\) and \(y_{q_4}\), if the quadratic coefficients of the quadratic variables is zero, then the number of quadratic Gauss sums reduces from \(3^2\) (indexed by \(y_{q_5}\) and \(y_{q_6}\)) to at most \(3\). This reduces the sum to \(3^5\) quadratic Gauss sums.

Otherwise, if the quadratic coefficients of the quadratic variables are not zero, then the Wigner function exhibits another property. The coefficient of the \(y_{q_6}^2\) term is only dependent on \(y_{q_2}\) of the cubic indexing intermediate variables. The coefficients of the \(y_{q_6}\) term is dependent on the \(y_{q_7}^2\) and \(y_{q_{11}}^2\) quadratic variables. Given a fixed \(y_{q_1}\), \(y_{q_2}\), \(y_{q_3}\), and \(y_{q_4}\) that does not set any of the quadratic coefficients of the quadratic variables to zero, it follows that for two values of \(y_{q_6}\) (and say \(y_{q_5} = 0\)), the quadratic coefficients of \(y_{q_7}^2\) and \(y_{q_{11}}^2\) are switched. Since these are separable quadratic coefficients, it follows that the Gauss sum at these two values of \(y_{q_6}\) has the same magnitude for all \(y_{q_5}\).

The Wigner function is real so the imaginary parts of these quadratic Gauss sums of equal magnitude must cancel out. Their complex conjugates lie across \(y_{q_1}\), \(y_{q_2}\), \(y_{q_3}\) or \(y_{q_4}\), but not \(y_{q_6}\). There are only three options for the phase for every \(y_{q_2}\) (corresponding to the three co-cycles) for the pairs of equal magnitude quadratic Gauss sums indexed by \(y_{q_6}\). For \(y_{q_2} = 0\), one of the options is for no imaginary part and so it follows that the quadratic Gauss sums in that sector that have equal magnitude must be real, or have the same imaginary part. For the two other sectors (\(y_{q_2} \ne 0\)), the same behavior holds since the only term that displaces a phase from being real is the sole \(y_{q_2}^3\) term, which is independent of \(y_{q_5}\) and \(y_{q_6}\).

This latter case produces \(3^4 \times 6 = 486\) quadratic Gauss sums, which is the worst case.

        This can be summarized by the following equation:
\begin{eqnarray}
  && \rho(\bsmc M_{C_{12}} (x_{p_1}, \ldots, x_{p_{12}}, x_{q_1}, \ldots, x_{q_{12}})) \nonumber\\
  &=& \sum_{\substack{y_{q_1},\, \ldots,\, y_{q_6}\\ \in \mathbb Z/ 3^2 \mathbb Z}} \exp \left[ \frac{2 \pi i}{9} \left( 7 y_{q_2}^3 + 8 x_{q_2}^3 \right) \right] \exp \left[ \frac{2 \pi i}{3} \Gamma_{12}(y_{q_1}, \ldots , y_{q_6}, \bs x_p, \bs x_q) \right] \mathcal A_{12}(\bs y_{q}, \bs x_p, \bs x_q)\\
  && \times \Bigg\{ \delta( \lor_{i=7}^{12}(\Sigma_{y_{q_i}} \land \Delta_{y_{q_i}}) ) + \delta( \neg (\lor_{i=7}^{12}(\Sigma_{y_{q_i}} \land \Delta_{y_{q_i}}))  ) \bigg[ \delta(\Sigma_{y_{q_{7}}}(0) - \Sigma_{y_{q_{11}}}(1))\big[2\delta(y_{q_6}) + \delta(y_{q_6}-2)\big] \nonumber\\
  && \qquad \qquad \qquad + \delta(\Sigma_{y_{q_{7}}}(1) - \Sigma_{y_{q_{11}}}(2))\big[2\delta(y_{q_6}-1) + \delta(y_{q_6})\big] + \delta(\Sigma_{y_{q_{7}}}(0) - \Sigma_{y_{q_{11}}}(2))\big[2\delta(y_{q_6}) + \delta(y_{q_6}-1)\big] \bigg] \Bigg\} \nonumber
\end{eqnarray}

The first term includes all cases where the quadratic coefficients of \(y_{q_7}\), \(\ldots\), \(y_{q_{12}}\) are zero along with their respective linear coefficients, as these produce plane waves that do not evaluate to zero. The second term includes all cases where all these quadratic coefficients are non-zero and \(y_{q_5}\) indexes the same quadratic Gauss sums for two values of \(y_{q_6}\) for some values of the the cubic variables.

The same simplification can be made for the final phase space variables \((\bs x_p, \bs x_q)\). In the worst-case, this produces \(\xi_{12} = 3^4 \times 6 = 486\) quadratic Gauss sums, which produces a tensor upper bound of \(\xi_{12}^{t/12} = 3^{\sim 0.469 t}\). 

Numerical examination of this Wigner function seems to indicate that this number can be lowered even further.

\section{Three Qutrit Monte Carlo Stabilizer Rank Numerical Search}
\label{app:numerics}

To numerically find the \(t=3\) qutrit \(T\) gate magic state stabilizer rank, we used the same algorithm as in~\cite{Bravyi16_2} but adapted for qutrits; we perform a random walk on the set of \(\chi\) stabilizer states and try to maximize the projection between the linear subspace they span and the \(k\)-tensored \(T\) gate magic state \(\phi\): \(F = ||\Pi \phi||\), where \(\Pi\) is the projector onto the linear subspace spanned by the stabilizer states. At each step, one of the stabilizer states \(\phi_i\) is randomly selected and a random Pauli operator is applied to it: \(\phi_i \rightarrow (I - P)(I - \omega P)\phi_i\), for \(\omega = \exp \frac{2 \pi i}{3}\). The new (renormalized) stabilizer state is accepted if it increases \(F\)'s value. It is rejected if \((I-P)(I-\omega P)\phi_i = 0\). Otherwise it is accepted with probability \(\exp \left[ -\beta( F - F') \right]\), where \(F\) and \(F'\) are the values of the projection before and after the step, respectively. The walk is stopped when \(F=1\), its maximum. We begin with a small \(\beta\) and ``anneal'' to a large final value.

\begin{figure}[ht]
  \includegraphics[scale=0.25]{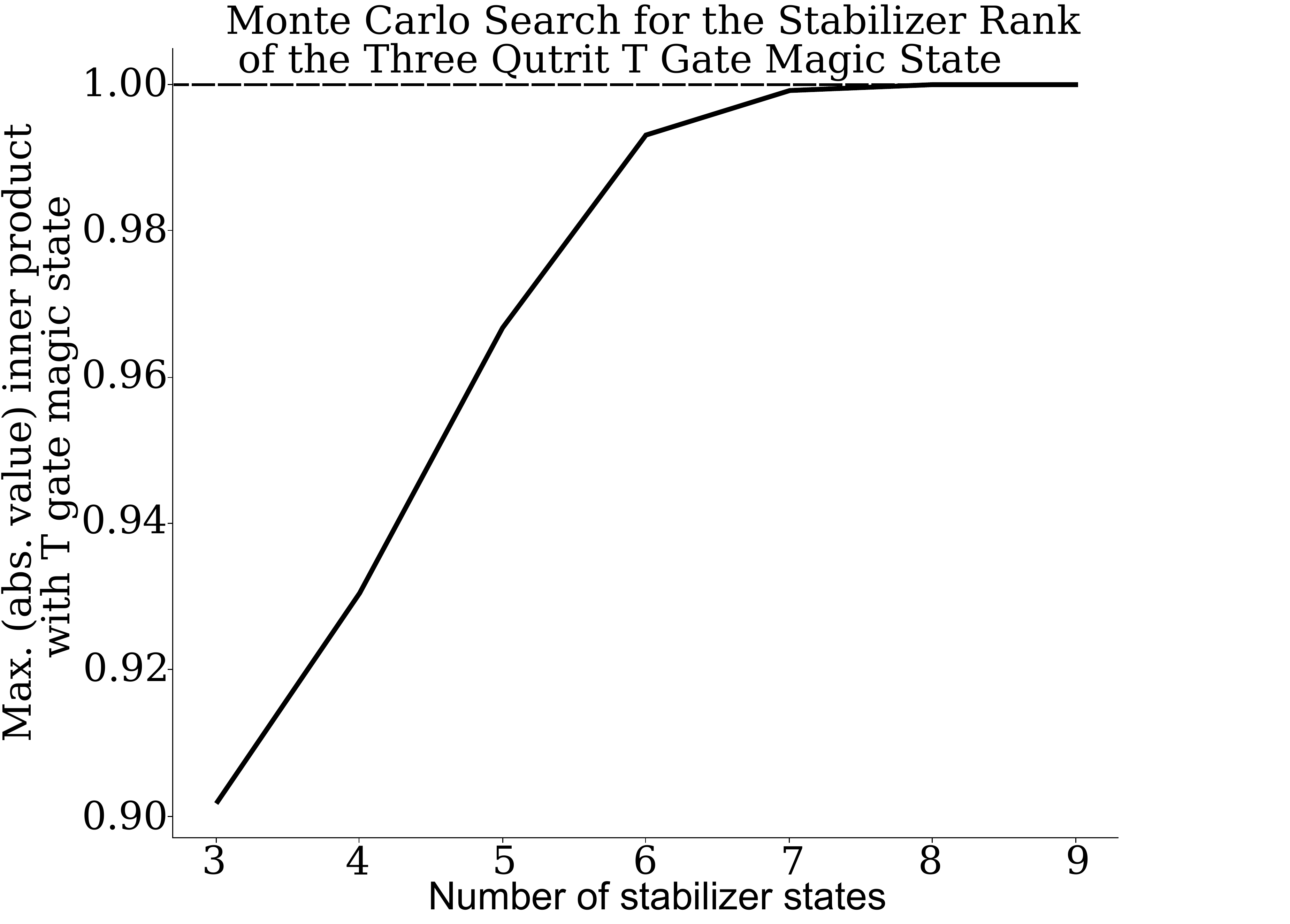}
  \caption{Three qutrit Monte Carlo stabilizer rank search.}
  \label{fig:threequtritMCstabranksearch}
\end{figure}

This approach produces results that are possibly not converged for the \(3\)-tensored qutrit \(T\) gate magic state at lower numbers of stabilizer states, since the search space is very large (comparable to \(t=7\) for qubits). The results are illustrated in Figure~\ref{fig:threequtritMCstabranksearch} and show a stabilizer decomposition upper bound of \(8\).

\end{document}